\documentclass[12pt]{article}

\usepackage[normalem]{ulem}
\usepackage[mathscr]{eucal}
\usepackage[version=3]{mhchem}
\usepackage{stmaryrd}

\usepackage{epsfig,latexsym,amsfonts,amsmath,amsthm,amssymb,amsbsy,multirow,slashed,wasysym,textcomp,dsfont,comment,mathtools,cancel,cite,datetime,appendix}
\usepackage{tocloft}
\usepackage{bbold}
\usepackage{tikz}
\usetikzlibrary{decorations.pathreplacing,calligraphy}
\usetikzlibrary{backgrounds}
\usetikzlibrary{patterns}
\usetikzlibrary{arrows.meta,calc}
\tikzstyle{bag} = [align=center]
\usetikzlibrary{decorations.pathmorphing}
\usepackage{enumitem}

\usepackage{hyperref}
\usepackage{subcaption}

\usepackage{array}
\usepackage{booktabs}

\newcommand{\badat}{\begin{alignedat}}
\newcommand{\eadat}{\end{alignedat}}
\newcommand{\mh}[1]{\hat{\mathcal{#1}}}

\def\be{\begin{equation}}
\def\ee{\end{equation}}
\def\ba{\begin{aligned}}
\def\ea{\end{aligned}}

\def\p{\partial}

\usepackage{xcolor}

\definecolor{dblue}{rgb}{0.2,0.50,0.80}

\def\N{\mathcal{N}}

\def\tr{\mathrm{tr}}

\def\bz{{\bar z}}

\newcommand{\Pbar}{\bar P}

\makeatletter
\newsavebox{\@brx}
\newcommand{\llangle}[1][]{\savebox{\@brx}{\(\m@th{#1\langle}\)}%
  \mathopen{\copy\@brx\kern-0.5\wd\@brx\usebox{\@brx}}}
\newcommand{\rrangle}[1][]{\savebox{\@brx}{\(\m@th{#1\rangle}\)}%
  \mathclose{\copy\@brx\kern-0.5\wd\@brx\usebox{\@brx}}}
\makeatother

\DeclareFontFamily{OT1}{pzc}{}
\DeclareFontShape{OT1}{pzc}{m}{it}{<-> s * [1.10] pzcmi7t}{}
\DeclareMathAlphabet{\mathpzc}{OT1}{pzc}{m}{it}

\definecolor{vert}{rgb}{0.1367 0.543 0.1367}

\usepackage[retainorgcmds]{IEEEtrantools}
\def\bea{\begin{IEEEeqnarray*}}
\def\eea{\end{IEEEeqnarray*}}

\renewcommand{\(}{\left(}
\renewcommand{\)}{\right)}

\newcommand{\beq}{\begin{equation}}
\newcommand{\eeq}{\end{equation}}
\newcommand{\beqa}{\begin{eqnarray}}
\newcommand{\eeqa}{\end{eqnarray}}

\definecolor{myblue}{RGB}{0, 122, 255}
\definecolor{mydarkblue}{RGB}{0, 98, 204}
\definecolor{myred}{RGB}{220,70,50}
\definecolor{mydarkred}{RGB}{176, 56, 40}
\tikzset{
  cyloutline/.style={line width=0.9pt, draw=black},
  inplane/.style={line width=0.9pt},
  vertline/.style={line width=0.9pt},
  guide/.style={line width=0.9pt, -{Stealth[length=3.5mm]}},
  guide2/.style={line width=0.9pt, {Stealth[length=3.5mm]}-{Stealth[length=3.5mm]}},
  dot/.style={circle, fill=black, inner sep=1.6pt, draw=none},
  labeltxt/.style={scale=1.05},
  every path/.style={line cap=round} 
}

\newcommand{\ColSep}{6.5}   
\newcommand{\RowSep}{6.0}   

\newcommand{\CylH}{4.6}     
\newcommand{\CylR}{1.65}    
\newcommand{\TopDepth}{0.55}

\newcommand{\SqSize}{3.6}
\newcommand{\CircR}{1.25}   

\tikzset{
  cutearr/.style={
    line width=0.8pt,
    -{Stealth[length=2.4mm,width=1.6mm]},
  },
}
\newcommand{\CuteGap}{0.14cm}   
\newcommand{\CuteSize}{0.3cm}  

\newcommand{\CuteGapWIDE}{1.1cm}
\newcommand{\CuteSizeWIDE}{0.5cm}

\newcommand{\CuteArrowsAt}[2][45]{%
  \foreach \k in {0,1,2,3}{%
    \pgfmathsetmacro{\ang}{#1 + 90*\k}%
    \draw[cutearr]
      ($ #2 + ({\CuteGap*cos(\ang)},{\CuteGap*sin(\ang)}) $)
        -- ++({\CuteSize*cos(\ang)},{\CuteSize*sin(\ang)});
  }%
}

\newcommand{\CuteArrowsWIDE}[2][45]{%
  \foreach \k in {0,1,2,3,4,5}{%
    \pgfmathsetmacro{\ang}{#1 + 60*\k}%
    \draw[cutearr]
      ($ #2 + ({\CuteGapWIDE*cos(\ang)},{\CuteGapWIDE*sin(\ang)}) $)
        -- ++({\CuteSizeWIDE*cos(\ang)},{\CuteSizeWIDE*sin(\ang)});
  }%
}

\newcommand{\CuteArrowsAtCYL}[2][]{%
  \pgfmathsetmacro{\ang}{90+3-30}%
    \draw[cutearr]
      ($ #2 + ({\CuteGap*cos(\ang)},{\CuteGap*sin(\ang)}) $)
        -- ++({\CuteSize*cos(\ang)},{\CuteSize*sin(\ang)});
  \pgfmathsetmacro{\ang}{90+3 + 30}%
    \draw[cutearr]
      ($ #2 + ({\CuteGap*cos(\ang)},{\CuteGap*sin(\ang)}) $)
        -- ++({\CuteSize*cos(\ang)},{\CuteSize*sin(\ang)});
  \pgfmathsetmacro{\ang}{270+3-30}%
    \draw[cutearr]
      ($ #2 + ({\CuteGap*cos(\ang)},{\CuteGap*sin(\ang)}) $)
        -- ++({\CuteSize*cos(\ang)},{\CuteSize*sin(\ang)});
  \pgfmathsetmacro{\ang}{270+3+30}%
    \draw[cutearr]
      ($ #2 + ({\CuteGap*cos(\ang)},{\CuteGap*sin(\ang)}) $)
        -- ++({\CuteSize*cos(\ang)},{\CuteSize*sin(\ang)});
}

\newcommand{\EllipseArrowLen}{0.5cm}   
\newcommand{\EllipseArrowLift}{-3 pt}     
\newcommand{\EllipseArrowSkipBack}{false} 

\newcommand{\UpArrowsOnEllipse}[5]{%
  \pgfmathtruncatemacro{\N}{#4}
  \pgfmathtruncatemacro{\Nminusone}{\N-1}
  \foreach \k in {0,...,\Nminusone}{%
    \pgfmathsetmacro{\ang}{#5 + 360*\k/\N}%
    \pgfmathsetmacro{\cx}{#2*cos(\ang)}%
    \pgfmathsetmacro{\cy}{#3*sin(\ang)}%
    \ifthenelse{\equal{\EllipseArrowSkipBack}{true}}
      {\pgfmathsetmacro{\doit}{(\cx>=0)?1:0}}
      {\pgfmathsetmacro{\doit}{1}}%
    \ifnum\doit=1
      \draw[cutearr]
        ($ #1 + (\cx,\cy) + (0,\EllipseArrowLift) $)
          -- ++(0,\EllipseArrowLen);
    \fi
  }%
}

\newcommand{\DrawCylinder}[4]{%
  \begin{scope}[shift={#1}]
    \draw[cyloutline] (-\CylR,-\CylH/2) -- (-\CylR,\CylH/2);
    \draw[cyloutline] ( \CylR,-\CylH/2) -- ( \CylR,\CylH/2);
    \draw[vertline, draw=#2] (.5,-\CylH/2+.52) -- (.5,\CylH/2+.52);
    \draw[inplane, draw=#3] (0,0) ellipse [x radius=\CylR, y radius=\TopDepth];
    \draw[cyloutline] (0,\CylH/2) ellipse [x radius=\CylR, y radius=\TopDepth];
    \draw[cyloutline] (0,-\CylH/2) ellipse [x radius=\CylR, y radius=\TopDepth];
    \draw[vertline, draw=#2] (-.5,-\CylH/2-.52) -- (-.5,\CylH/2-.52);
    \ifx#4t\relax
      \node[dot] at (\CylR-.25,-.3) {};
    \fi
  \end{scope}
}

\newcommand{\DrawSquarePanel}[5]{%
  \begin{scope}[shift={#1}]
    \pgfmathsetmacro{\h}{\SqSize/2}
    \draw[cyloutline] (-\h,-\h) rectangle (\h,\h);
    \draw[vertline, draw=#2] (0,-\h) -- (0,\h);
    \draw[inplane, draw=#3] (0,0) circle (\CircR);
    \ifx#4t\relax
      \node[dot] at (\CircR,0) {};
    \fi
    \ifx#5t\relax
      \node[dot] at (0,0) {};
    \fi
  \end{scope}
}

\numberwithin{equation}{section} 

\begin{document}

 \begin{titlepage}
  \thispagestyle{empty}
  \begin{flushright}
  \end{flushright}
  \bigskip
  \begin{center}

        \baselineskip=13pt {
        \Large{\bf { Swing Surfaces in AdS/CFT}
       }}

      \vskip1cm 

   \centerline{ 
   {Jacqueline Caminiti}${}^{1,{}2}$ and {Sabrina Pasterski}\,${}^1$
}

\bigskip\bigskip

 \centerline{\em${}^1$ 
\it Perimeter Institute for Theoretical Physics, Waterloo, ON N2L 2Y5, Canada}

\bigskip

\centerline{\em${}^2$ 
\it Department of Physics \& Astronomy, University of Waterloo, 
	Waterloo, ON N2L 3G1, Canada}

\bigskip\bigskip

\end{center}

\begin{abstract}

\noindent 
We study holographic entanglement entropy in the flat limit of AdS$_3$/CFT$_2$, and clarify the relation between Ryu-Takayanagi surfaces and the swing surfaces proposed for flat-space holography. We show that the ordinary RT formula does not admit a well-defined flat-space limit and connect this to subtleties of the state-operator map in the flat limit. We then construct an AdS$_3$ uplift of the swing prescription and develop its corresponding information-theoretic interpretation through two complementary approaches. First, we show that the swing surface length computes the mutual information between suitably chosen nearly null boundary intervals in a special class of pure CFT states. The gravity interpretation of this result involves scattering processes which take place deep inside of the AdS$_3$ bulk. Second, we reduce the problem of finding an information-theoretic interpretation for swing surfaces to the more general problem of finding an information-theoretic interpretation of black hole inner horizons in
holography.

\end{abstract}

\end{titlepage}

\tableofcontents

\vspace{1em}
\noindent

\section{Introduction}
\label{sec:intro}

The holographic principle provides a powerful window into quantum gravity in asymptotically anti-de Sitter (aAdS) spacetimes, via the well-established AdS/CFT correspondence.
Meanwhile, one expects from black hole thermodynamics that the holographic principle is more widely applicable than the aAdS setting.  One is then compelled to ask:
What is the holographic description of quantum gravity in asymptotically flat spacetimes?
And can this description be obtained from a controlled limit  within the framework of AdS/CFT?

This limit is not as straightforward as one might hope. The flat limit of the AdS factor in usual top-down constructions is incompatible with keeping the compactified directions small. Even bottom-up approaches, which would seem more agnostic to the stringy origins of the duality, run into trouble: the naive flat-space limit of the bulk forces the relevant part of the timelike boundary of AdS to degenerate into a null surface.
We are left with a much more exotic Carrollian theory, and it is not always clear which states, operators, symmetry representations, and information-theoretic observables of the parent CFT remain well-defined in this limit. In this work, we investigate this problem in the  setting of AdS$_3$/CFT$_2$ and its flat limit.

The case of three-dimensional bulks is particularly tractable because the asymptotic symmetry group is infinite-dimensional.
In AdS$_3$/CFT$_2$, this Virasoro symmetry strongly constrains boundary observables. 
Moreover,  
in many circumstances,
we understand how the BMS algebra of the Carrollian theory arises from  Vir$\times$ Vir in the flat limit. 
The seminal works of Bagchi and collaborators \cite{Bagchi:2010zz,Bagchi:2012cy} used this BMS$_3$ contraction of the Virasoro symmetry to bootstrap observables in the putative Carrollian boundary theory \cite{Bagchi:2017cpu}, including subregion entanglement entropies \cite{Bagchi:2014iea}, our focus in this work. 
In AdS/CFT, the leading semiclassical entropy of a boundary interval is computed by the length of its Ryu-Takayanagi surface \cite{Ryu:2006bv,Hubeny:2007xt,Lewkowycz:2013nqa}. 
In flat-space holography, however, a strikingly different geometric prescription has emerged. 
In~\cite{Jiang:2017ecm}, Wei Song and collaborators showed that the entropy computed in \cite{Bagchi:2014iea} is reproduced by the length of a bulk extremal curve known as a \emph{swing surface}.
Notably, rather than being anchored at the asymptotic boundary, the swing surface is extremized between two null curves --- called the \emph{swing ropes} \cite{Apolo:2020bld} --- which emanate from the endpoints of the boundary interval and extend deep into the bulk.

The goal of this work is to situate the swing proposal within AdS/CFT.
Our guiding principle is that any meaningful information-theoretic quantity in flat-space holography should lift to a meaningful quantity in AdS/CFT,
whether it coincides with the standard entanglement entropy or not.
From this perspective, the central question is not simply whether an RT surface becomes a swing surface in the flat-space limit, but rather what information-theoretic structure in AdS/CFT gives rise to the swing formula. This perspective is complementary to the AdS/BCFT approach being developed in Refs.~\cite{Hao:2025ocu,SivilottiMyers:WIP},
the timelike entanglement and pseudoentropy approaches of \cite{Wen:2024yny,Fareghbal:2024lqa},
as well as the top-down approach involving D$p$ branes in \cite{Jorstad:2026jlg}.

We begin by examining the flat-space limit of the ordinary RT formula. 
We find that ordinary RT surfaces do not admit a well-defined flat-space limit: their areas diverge, and the surfaces fail to localize within any approximately flat region in the bulk.
We then show that this divergence matches a boundary entropy computation using twist operators in the Carrollian limit.
In particular, we show that the RT divergence can be 
traced to a mismatch between the central charges for the contracted BMS algebra under which the twist operators are BMS primaries,
versus the contracted BMS algebra used in asymptotic symmetry analyses. The only difference between these two cases is a change of frame in the parent CFT, and hence this appears to point to a broader lesson whereby the limiting BMS theory depends heavily on an original choice of conformal frame. 
This insight then allows us to construct a natural uplift of the swing prescription to  AdS$_3$, and we then pursue two complementary approaches to understand its information-theoretic interpretation.

First, we show that in the CFT vacuum state,
the swing surface computes the mutual information between a pair of suitably chosen, nearly null boundary intervals:
\begin{equation}
\frac{\operatorname{Length}(\gamma_{\mathrm{swing}})}{4G}
=
\frac{1}{2}I(V_1:V_2).
\label{iden}
\end{equation}
This identifies the swing prescription with a familiar, positive, and UV-finite observable of the parent CFT.
We then demonstrate that this identification admits a direct bulk interpretation in terms of scattering processes occurring deep inside the AdS$_3$ geometry, using the holographic scattering formalism of~\cite{May:2019odp}.
From this perspective,
the null swing ropes in the swing prescription naturally arise from the nearly null kinematics of the intervals entering the mutual information.

While this construction gives an interesting template for how an ordinary information-theoretic quantity in AdS/CFT can isolate flat-space physics,
the precise identification in \eqref{iden} is restricted to the vacuum state,
together with a special class of pure states for which the entropy factorizes into left- and right-moving pieces.
This then motivates us to pursue a second approach to interpreting swing surfaces, 
based on a refinement of the original work of \cite{Jiang:2017ecm}.

In this approach, our starting point is the conformal map of Casini-Huerta-Myers \cite{Casini:2011kv}, which relates the vacuum reduced density matrix of a boundary interval to an ordinary thermal state. 
Following~\cite{Jiang:2017ecm} (and more recently~\cite{Fareghbal:2020dtq}), we modify the CHM construction by endowing the thermal state with a nonzero chemical potential, so that the bulk dual is described by a spacetime with both an outer and an inner horizon. 
Upon returning to the original conformal frame, the outer horizon maps to the usual RT surface, while the inner horizon maps to the swing surface; moreover, the null swing ropes emerge naturally from a choice of UV regulator in the boundary CFT. 
We can write this as the following identification:
\begin{equation}
\frac{\operatorname{Length}(\gamma_{\mathrm{swing}})}{4G}
=
\frac{\operatorname{Length}(\gamma_{\mathrm{in}})}{4G}\,.
\label{eq:intro-inner-horizon}
\end{equation}
The upshot is that this identification reduces the problem of interpreting swing surfaces to the more general question of identifying the information-theoretic meaning of inner horizon areas in holography.
Motivated by this, we outline how the Cardy formula in $2d$ CFT appears to receive a subleading saddle contribution from the inner horizon area which becomes dominant in the flat-space limit. 
In this limit, the outer horizon area diverges, similarly to the divergent RT surfaces addressed in section~\ref{sec:RTflat}, but now the actual BTZ solution has finite-area horizons, and this lets us avoid the subtlety of prescribing radial ropes.

Taken together, our results demonstrate that the flat-space limit does not simply deform the RT prescription into the swing prescription; 
rather, it selects a distinct information-theoretic structure already present in AdS/CFT, which opens new avenues for investigation within AdS/CFT itself as well as inner horizon physics.

This paper is organized as follows. In section~\ref{sec:basics}, we review the flat-space limit of asymptotically AdS$_3$ geometries and its realization as a Carrollian limit of the boundary CFT$_2$. In section~\ref{sec:RTflat}, we analyze the failure of the RT prescription in this limit and describe the associated central charge mismatch of the BMS algebra arising from a change of frame.
In section~\ref{sec:swing}, we construct swing surfaces in AdS$_3$ and compare their lengths with an uplifted boundary formula. 
In section~\ref{sec:lift}, we identify swing surfaces with mutual informations of nearly null regions and develop their bulk-scattering interpretation. In section~\ref{sec:innerhor}, we relate swing surfaces to inner horizons through a modified CHM construction, and connect inner horizons to a subleading Cardy saddle.
We conclude with open questions concerning flat-space observables, the covariance of the swing ropes,  
and connections to holographic scattering and timelike entanglement entropy.
The appendices hold technical results generalizing derivations in the main text.

\section{Flat limit as a Carrollian limit}
\label{sec:basics}

In this section, we review  how the flat-space limit of asymptotically AdS$_{3}$ spacetimes is realized as a Carrollian limit of the dual boundary theory. We start with the specific coordinate limits in section~\ref{sec:lim} before discussing the contracted symmetry algebras in section~\ref{sec:asg}.

\subsection{Coordinate limits}\label{sec:lim}
To build intuition, we begin with a simple  example: the flat-space limit of vacuum AdS$_{3}$.
Written in global coordinates,
the vacuum AdS$_{3}$ metric reads 
\be\label{ads}
ds^2=-\left(\frac{r^2}{\ell^2}+1\right)\ell^2d\tau^2+\frac{1}{\frac{r^2}{\ell^2}+1}dr^2+r^2 d\phi^2\,,
\ee
with $r>0$, $\phi\sim \phi+2\pi$, and $\tau\in \mathbb{R}$ a dimensionless time coordinate.
Here $\ell$ is the AdS length scale, related to the Ricci curvature scale by $R=-6/\ell^2$.
Taking the flat-space limit means sending $R$ to zero, or equivalently, taking $\ell$  to infinity, with respect to all other scales in the problem.

It is clear how this should work in \eqref{ads}: in the regime $\ell \gg r,|t|$ for~$\ell\tau\equiv  t$,
\eqref{ads} reduces to the metric of three-dimensional Minkowski spacetime,
\be\label{flat}
ds^2=-dt^2+dr^2+r^2 d\phi^2\,,
\ee
in polar coordinates.
This is our first example of a flat limit,
and in this case, 
it is simply the statement that global AdS looks like Minkowski spacetime when one zooms into a tiny sub-AdS scale region. 

\begin{figure}
    \centering
    \includegraphics[width=.5\linewidth,trim= 0 0 3em 1em]
    {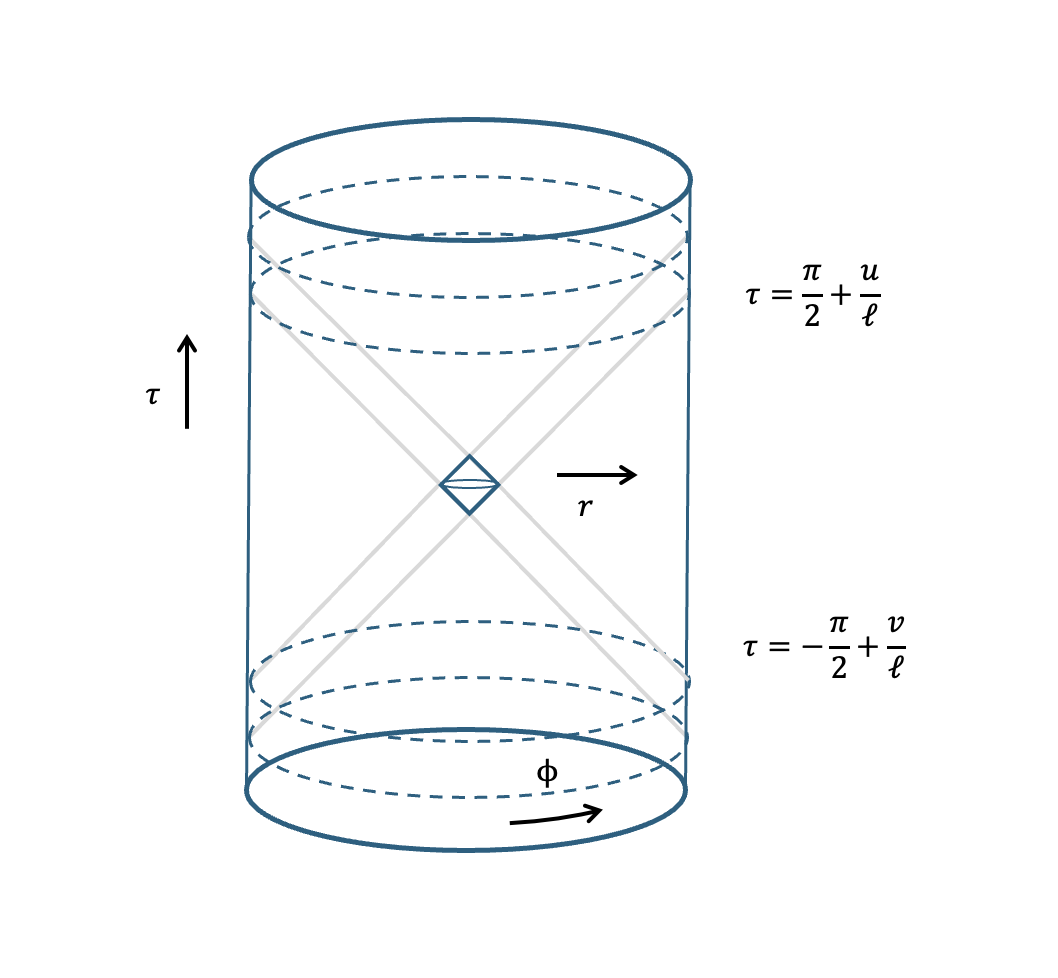}
    \caption{This figure shows the conformal compactification of global AdS$_3$. 
    The approximately flat region with $r\ll \ell$ 
    and $|\tau| \ll 1$
    is schematically drawn as the region bounded by the blue diamond deep in the middle of AdS.
    We see that it can send null rays up to the Carrollian boundary time band with $|\tau-\frac{\pi}{2}|\ll 1$.
   }
    \label{fig:bulk-point}
\end{figure}

This sub-AdS scale patch 
is far away from the conformal boundary, but lightrays emanating from it will eventually hit the AdS boundary. 
With the change of coordinates \cite{Barnich:2012aw} 
\be\label{eq:jumper}
\tau=\frac{u}{\ell}+\arctan\frac{r}{\ell},
\ee
the metric~\eqref{ads} takes the Bondi form:
\be\label{uads}
ds^2=-\left(\frac{r^2}{\ell^2}+1\right)du^2-2dudr+r^2d\phi^2\,,
\ee
and we can more readily examine the boundary region reached by null geodesics.
In these coordinates, rays with $u$ and $\phi$ fixed are outgoing null rays.
In particular, traveling from $r=0$ to $r\to\infty$ along the null ray with $u=0$ causes $\tau$ to jump by $\pi /2$. 
This gives rise to the picture in figure \ref{fig:bulk-point},
in which Minkowski space lives deep inside of the AdS bulk near $\tau=0$,
but can send null rays to a band of the AdS boundary centered at $\tau=\frac{\pi}{2}$.

\begin{figure}
    \centering
    \includegraphics[width=.85\linewidth,trim= 0 3em 0 0]
    {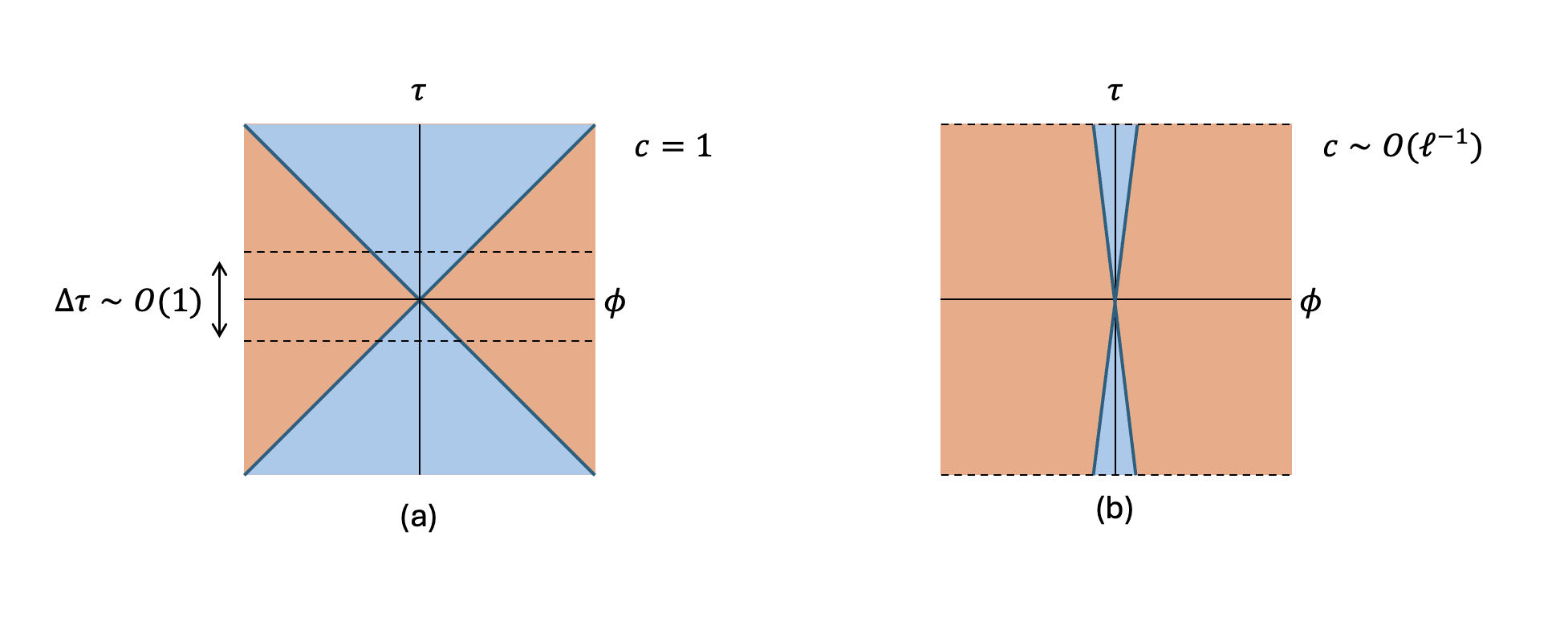}
    \caption{From the boundary perspective, the flat limit takes us from the relativistic (a) to ultrarelativistic (b) theory. 
    In particular, zooming into short time scales $\Delta \tau\ll 1$ stretches the lightcone into a Carrollian regime.
    }
    \label{fig:carroll-limit}
\end{figure}

Let us now understand how the flat-space limit acts in the dual CFT.
We will think of the dual CFT as living on the conformal boundary of the spacetime described in~\eqref{ads} and~\eqref{uads}.
Its metric can be obtained by considering large, fixed values of $r$ and stripping off a conformal factor
\be
ds^2\simeq r^2\left[-d\tau^2+d\phi^2\right]\simeq r^2\left[-\frac{1}{\ell^2}du^2+d\phi^2\right]\,.
\label{bdymetric}
\ee
That is, $ds^2=r^2 ds^2_\p$ identifies the boundary metric.\footnote{Since $r$ has units of length, stripping off $r$ technically requires introducing a new, fixed scale $r_0$ via $ds^2=\frac{r^2}{r_0^2} ds^2_\p$. For ease of notation, we suppress $r_0$ in what follows, which explains why some expressions below appear dimensionally unbalanced.
}
We see that 
the $\ell\to\infty$ limit corresponds to zooming into a time band on the conformal boundary with small dimensionless extent $\Delta \tau \sim \Delta u/\ell$.
Equivalently, since $\frac{1}{\ell^2}$ multiplies the time component in \eqref{bdymetric}, $\frac{1}{\ell}$ plays the role of the speed of light in the boundary. When we take $\ell$ to be much larger than all other scales in the problem, the limit $c\sim\frac{1}{\ell}\rightarrow 0$  gives us a degenerate metric, in which the $u$ direction, which was timelike in \eqref{uads},
has now become a null direction  of the boundary theory:
\be
ds^2_\p=0\cdot du^2+d\phi^2.
\label{carbdymetric}
\ee 
This type of degenerate metric is called a Carrollian metric, and the $c\to0$ limit which takes \eqref{bdymetric} to \eqref{carbdymetric} is called a Carrollian or ultrarelativistic limit (see figure~\ref{fig:carroll-limit}), to be contrasted with the Galilean or non-relativistic $c\to\infty$ limit.
Following \cite{Heemskerk:2009pn,Hijano:2019qmi} (see also~\cite{PipolodeGioia:2022exe,deGioia:2024yne}), one can argue that the small Carrollian time band near $\tau=\frac{\pi}{2}+\frac{u}{\ell}$ is associated with future null infinity while the small Carrollian time band near $\tau=-\frac{\pi}{2}+\frac{v}{\ell}$ is associated with past null infinity.

In summary, the above example illustrates a direct geometric connection between bulk flat-space limits and  boundary Carrollian limits.
While we have restricted our attention to the global AdS$_3$ spacetime in the bulk, 
one can also generalize the above argument to global AdS spacetimes in any dimension, and this is referred to in the literature as the bulk point limit \cite{Heemskerk:2009pn,Maldacena:2015iua,Hijano:2019qmi,PipolodeGioia:2022exe,deGioia:2024yne}.
One can also generalize the above argument to any vacuum solution of Einstein's equations with negative cosmological constant in $d+1=3$ spacetime dimensions \cite{Bagchi:2012xr,Barnich:2010eb,Barnich:2012aw}.
For instance, one can consider the AdS$_3$ spinning defect geometries \cite{Deser:1983tn}.
These are solutions of AdS$_3$ gravity with ADM mass $-1/(8G)<M<0$ and spin $J$.
If one repackages $M$ and $J$ via the following parameters  \cite{Li:2024rma}:
\begin{equation}
    b_\pm = \sqrt{2G \ell}\left(\sqrt{\ell |M| + J}\pm  \sqrt{\ell |M| -J}\right)\,,
    \label{eq:M-J-from-rpm}
\end{equation}
then the AdS$_3$  spinning defect geometries can be described as follows:
\begin{equation}
ds^2 = 
-
\frac{(r^2+b_+^2)(r^2+b_-^2)}{\ell^2 r^2} dt^2
+
\frac{\ell^2 r^2 dr^2}{(r^2+b_+^2)(r^2+b_-^2)} 
+ r^2 \left(d\phi - \frac{b_+ b_-}{\ell r^2} dt\right)^2\,.
\label{eq:defect-metric-bpm}
\end{equation}
To take the flat limit of this geometry, one can send $\ell\to\infty$ with $M$ and $J$ held fixed (and with $r,|t|\ll \ell$).
In this limit, $b_+$ diverges with $\ell$, so we define a new quantity
\begin{equation}
    \hat{b}_+=b_+/\ell \,.
\end{equation}
With this, one recovers the analogous flat-space spinning defect geometry in the bulk \cite{Deser:1983tn}:
\begin{equation}
ds^2 = 
-
\frac{\hat{b}_+^2(r^2+b_-^2)}{r^2} dt^2
+
\frac{r^2 dr^2}{\hat{b}_+^2(r^2+b_-^2)} 
+ r^2 \left(d\phi - \frac{\hat{b}_+ b_-}{r^2} dt\right)^2\,.
\label{eq:defect-metric-bpm-flat}
\end{equation}
This procedure generalizes \cite{Barnich:2012aw} from the specific geometries considered here to more general Ba\~nados geometries~\cite{Banados:1998gg}.
We thus have a coordinate limit that takes us from aAdS$_3$ metrics to AFS$_3$ metrics.
Meanwhile, the conformal boundary of \eqref{eq:defect-metric-bpm} has the same structure as in \eqref{bdymetric}, so we conclude that sending $\ell\to\infty$ again induces a Carrollian limit on the boundary.

\subsection{Asymptotic symmetry group}\label{sec:asg}

In three dimensions, the flat-space limit framework extends beyond the coordinate limit on solutions
to the full asymptotic symmetry group. 
Recall that 
one of the key entries in the AdS$_3$/CFT$_2$ dictionary is the nontrivial matching between the asymptotic symmetry algebra of AdS$_3$ and the conformal symmetry algebra of the dual two-dimensional CFT.
Namely, both of these symmetry algebras are described by two commuting Virasoro algebras \cite{Brown:1986nw}:
\begin{equation}
\begin{aligned}
    [L_n, L_m]
    &= 
    (n-m)L_{n+m}
    + 
    \frac{c}{12}n(n^2-1) \delta_{n+m}\,,\\
    [\bar{L}_n, \bar{L}_m]
    &= 
    (n-m)\bar{L}_{n+m}
    + 
    \frac{\bar{c}}{12}n(n^2-1) \delta_{n+m}\,,
    \\
    [L_n,\bar{L}_m]&=0\,.
\end{aligned}
\label{virasoro}
\end{equation}
In the CFT$_2$, $L_n$ and $\bar{L}_n$ generate conformal symmetries, and 
$c$ and $\bar{c}$ describe the left- and right-moving  central charges of the theory.
Meanwhile, in the bulk AdS$_3$ spacetime, $L_n$ and $\bar{L}_n$ are interpreted as generating asymptotic symmetries of the gravity theory, and $c$ and $\bar{c}$ characterize central extensions of the bulk's asymptotic symmetry algebra, computed using covariant phase space methods.
These take on the following values \cite{Brown:1986nw}
\begin{equation}
    c=\bar c=\frac{3\ell}{2G}
    \label{einsteinc}
\end{equation}
for Einstein gravity, but may vary depending on the particular theory of gravity living in the bulk. In particular, if we augment  the Einstein-Hilbert term in the bulk gravity action 
by the following gravitational Chern-Simons term:
\begin{equation}
\begin{aligned}
S_{\mathrm{TMG}}= & \frac{1}{16 \pi G} \int d^3 x \sqrt{-g}\left(R+\frac{2}{\ell^2}\right) \\
& +\frac{1}{32 \pi G \mu} \int d^3 x \sqrt{-g} \epsilon^{\lambda \mu \nu}\, \Gamma_{\lambda \sigma}^\rho\left(\partial_\mu \Gamma_{\rho \nu}^\sigma+\frac{2}{3} \Gamma_{\mu \tau}^\sigma \Gamma_{\nu \rho}^\tau\right) \,,
\end{aligned}
\label{tmg}
\end{equation}
we instead find \cite{Kraus:2005zm}:
\begin{equation}
    c=\frac{3\ell}{2G}\left(1+\frac{1}{\mu \ell}\right)
    \quad {\rm and}\quad
    \bar{c}=\frac{3\ell}{2G}\left(1-\frac{1}{\mu \ell}\right)\,.
    \label{cftcharge}
\end{equation}
The Chern-Simons term in \eqref{tmg} breaks parity in the bulk,
and leads to a mismatch between left and right central charges in the boundary theory.
The resulting theory goes by the name of topologically massive gravity (TMG) \cite{Deser:1982vy,Deser:1981wh}, because the topological term introduces a propagating spin-two mode with mass $\mu$. 
Note that the classical solutions of TMG include all solutions of the corresponding Einstein theory, which is recovered in full in the $\mu\to\infty$ limit.

Meanwhile, the asymptotic symmetry algebra of asymptotically flat spacetimes (AFS) is a BMS algebra. In three dimensions, this takes the form \cite{Barnich:2006av,Bagchi:2010zz,Bagchi:2012cy,Duval:2014uva}
\begin{equation}
\begin{aligned}
    [{\cal L}_n, {\cal L}_m]
    &=(n-m) {\cal L}_{n+m} + \frac{c_L}{12} n(n^2-1)\delta_{n+m}\\
    [{\cal L}_n, {\cal M}_m]
    &= (n-m) {\cal M}_{n+m} + \frac{c_M}{12} n(n^2-1)\delta_{n+m}
    \\
    [{\cal M}_n, {\cal M}_m]
    &= 0\,,
\end{aligned}
\label{bms3}
\end{equation}
where ${\cal M}_m$ are the supertranslations and ${\cal L}_m$ are superrotations. 
In the usual flat limit framework, one starts from the following combinations of Virasoro generators~\cite{Barnich:2012aw}
\be\label{bms}
  {\cal L}_n = L_n - \bar{L}_{-n},~~
         {\cal M}_n = \lim_{\ell\rightarrow \infty}\frac{1}{\ell}\left(L_n + \bar{L}_{-n} \right)
 \ee
and an In\"on\"u-Wigner contraction \cite{Inonu:1953sp} then recovers the algebra~\eqref{bms3} with central charges 
 \be\label{cstate}
  c_L = \lim_{\ell\rightarrow \infty} c-\bar{c}=\frac{3}{\mu G},~~
        c_M = \lim_{\ell\rightarrow \infty}  \frac{1}{\ell}\left(c+\bar{c}\right)=\frac{3}{G}\,,
\ee
agreeing with the asymptotic symmetry analyses of \cite{Barnich:2006av,Bagchi:2012yk}.
Note that the central charges $c_L$ and $c_M$ have well-defined flat limits thanks to the precise choice of relative signs in~\eqref{bms}-\eqref{cstate}; in particular, both $c$ and $\bar{c}$ individually diverge as $\ell\to\infty$.
The flat-space dictionary asserts that the Carrollian theory should enjoy a BMS$_3$ symmetry with central extensions \eqref{cstate}.

As we explore the difference between the usual RT surface and the swing proposals in the following sections, we will encounter many subtleties in this simple story.
In particular, we will find that \eqref{bms} is not the only BMS subalgebra needed to describe physics in the flat limit, and we will find associated ambiguities in the choices of central charges and what representations local operators live in.
Ultimately, we will unify these issues through a careful consideration of how the flat limit interplays with the CFT state-operator map.

\begin{figure}
    \centering
    \includegraphics[width=.8\linewidth,trim= 0 3em 0 0]{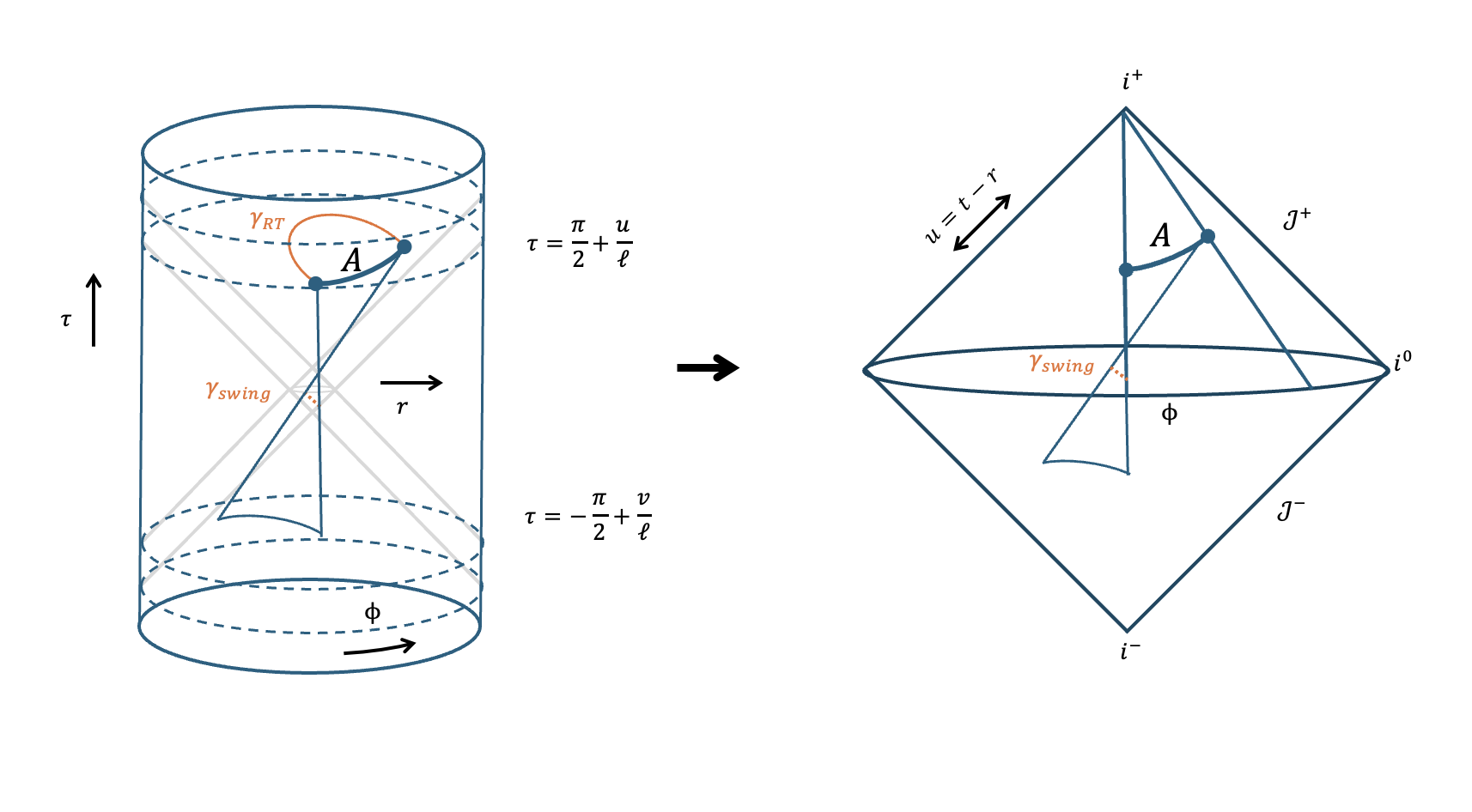}
    \caption{AdS (left) and flat-space limit (right) of a boundary subregion and its corresponding RT and swing surfaces. As we argue below, the RT surface becomes ill-defined in the flat limit.
    }
    \label{fig:flatlimit}
\end{figure}

\section{RT surfaces in the flat-space limit}
\label{sec:RTflat}

In this section, we show that the RT formula is not well-defined in the  flat limit, from both bulk and boundary perspectives, when using the flat limit procedure introduced in section \ref{sec:basics}.

\subsection{Bulk perspective}
\label{sec:bulk-persp}

To begin,
we will focus on the global AdS$_3$ spacetime in the bulk \eqref{ads},
dual to the vacuum state of a CFT$_2$ on the boundary cylinder. 
In order to compute the entanglement entropy of an interval $A$ with extent $(\Delta \tau,\Delta \phi)$, 
the RT formula\footnote{We use \textit{RT surface} as shorthand for the covariant HRT surface when the interval is not at constant time.}  instructs us to find the minimal extremal surface $\gamma_{\textrm{RT}}$ in the bulk which is homologous to the boundary interval $A$.
The relevant RT surface in global AdS$_3$ can be parameterized as follows:
\begin{equation}
\begin{aligned}
    \tau(\phi)-\tau_0 &=
     \tan^{-1} \left[\frac{\tan \frac{\Delta \tau}{2}}{\tan \frac{\Delta \phi}{2}}\tan(\phi-\phi_0)\right]
    \\
    r(\phi)^2 &= \frac{\ell^2 \sin^2 \Delta\phi}{\left(\cos {\Delta \tau} - \cos \Delta \phi\right)\left(\cos(2(\phi-\phi_0)) - \cos \Delta \phi\right)}\,.
\end{aligned}
\label{geod-og}
\end{equation}
Placing a UV cutoff at  $r=\ell/\varepsilon$, the intervening length is
\begin{equation}
    \textrm{Length}(\gamma_{\textrm{RT}}) 
    =
   \ell \log \left(\frac{2}{\varepsilon} \sin\frac{\Delta y^+}{2}\right)+
   \ell
\log \left(\frac{2}{\varepsilon} \sin\frac{\Delta y^-}{2}\right)\,,
\label{eq:length}
\end{equation}
where we have introduced the null coordinates 
\begin{equation}
    y^{\pm} = \phi \pm \tau
    \label{ypm}
\end{equation}
on the boundary cylinder. 
The RT formula in Einstein gravity says that the entanglement entropy of the boundary interval $A$ is given by $S(A) = \textrm{Length}(\gamma_{\textrm{RT}})/4G$. Indeed, plugging in \eqref{eq:length} reproduces the expected boundary formula 
\begin{equation}
    S(A)
    =
   \frac{c}{6} \log \left(\frac{2}{\varepsilon} \sin\frac{\Delta y^+}{2}\right)+
   \frac{\bar{c}}{6} 
\log \left(\frac{2}{\varepsilon} \sin\frac{\Delta y^-}{2}\right)\,
\label{SCFT}
\end{equation}
for the entanglement entropy of $A$, upon substituting the Einstein central charges $c=\bar c=\frac{3\ell}{2G}$ from~\eqref{einsteinc}. 
In topologically massive gravity, the RT formula involves an extra term measuring the twist of a normal frame along the bulk curve; using this, \cite{Castro:2014tta} demonstrated a bulk match to the $c\neq\bar c$ entropy~\eqref{SCFT}.

Our goal is to analyze~\eqref{eq:length}, \eqref{SCFT} 
in the flat-space limit.
This is straightforward once we take into account the coordinate scaling from section~\ref{sec:lim} relevant to the flat-space limit. In particular, we must scale the boundary-anchored points to stay in the band that becomes null infinity in the $\ell\rightarrow\infty$ limit. We do this by substituting
\begin{equation}\label{eq:null}
    y^{\pm} = \phi \pm u/\ell\,,
\end{equation}
and choosing intervals for which $\Delta u$ is held fixed as $\ell$ is sent to infinity.
Note that we have implicitly re-centered time so that the Carrollian time band is now centered at $\tau=0$ instead of $\tau=\pi/2$.

Having chosen an appropriate boundary interval to study, we may now examine  the associated RT surface in the $\ell\to\infty$ limit.  For generic endpoints, the geodesic in~\eqref{geod-og}  hangs down into the bulk to a minimal radius $r_*$:
\begin{equation}
    r_*^2= \ell^2\frac{1+\cos \Delta \phi}{\cos \Delta \tau - \cos \Delta \phi}\,.
    \label{eq:rturn}
\end{equation}
As we go to configurations with parametrically small $\Delta\tau$, the turning point becomes
\begin{equation}
    r_*= \ell\left|\cot \!\frac{\Delta\phi}{2}\right|\,.
    \label{eq:rturn2}
\end{equation}
We see that unless the angular sizes are tuned to be parametrically close to $\Delta\phi=\pi$, the geodesics will not probe a sub-AdS scale region in the center of AdS (i.e. $r\ll \ell$); see also \cite{Jorstad:2026jlg,Ghosh:2023kuy}. 
Even then, such geodesics are probing a sub-AdS scale region separated by $\pi/2$ AdS units 
in the time direction from the one of interest (see left-hand side of figure~\ref{fig:flatlimit}).

This misalignment is our first hint that the flat limit of the RT surface is ill-defined.
More decisively, one can simply study the length of the RT surface in the flat limit.
Substituting~\eqref{eq:null} into \eqref{eq:length} and taking the flat limit, it reads
\begin{equation}\label{bmsee2holo}
\textrm{Length}(\gamma_{\textrm{RT}})  \simeq  2\ell \ln \left(\frac{2}{\varepsilon} \sin \frac{\Delta \phi}{2}\right)+ \mathcal{O}(1/\ell), 
\end{equation}
which diverges as $\ell\to\infty$. 
For later reference, we note that for $c\ne \bar{c}$ ($\mu<\infty$), part of the $\mathcal{O}(1/\ell)$ piece of the associated entropy  reads
\begin{equation}
    \frac{c-\bar{c}}{12\ell }\Delta u \cot \frac{\Delta \phi}{2}
    =
    \frac{1}{4\ell\mu G}\Delta u \cot \frac{\Delta \phi}{2}
    \,.
    \label{bmsee2holo_subleading}
\end{equation}
A similar conclusion holds for more general asymptotically AdS spacetimes.
For example, in the AdS$_3$ spinning  defect geometry \eqref{eq:defect-metric-bpm}, we have 
\begin{equation}
\begin{aligned}
  & \textrm{Length}(\gamma_{\textrm{RT}}) \\
     &=\text{min}
    \begin{cases}
       \ell\log \left[\frac{2 \ell}{\varepsilon(b_++b_-)}\sin\left(\frac{b_++b_-}{2 \ell}\Delta y^+\right)\right]
    +
    \ell\log \left[ \frac{2\ell}{\varepsilon(b_+-b_-)}\sin\left(\frac{b_+ -b_-}{2 \ell}\Delta y^-\right)\right]\\
       \ell\log \left[\frac{2 \ell}{\varepsilon(b_++b_-)}\sin\left(\frac{b_++b_-}{2 \ell}(2\pi-\Delta y^+)\right)\right]
    +
    \ell\log \left[ \frac{2\ell}{\varepsilon(b_+-b_-)}\sin\left(\frac{b_+ -b_-}{2 \ell}(2\pi-\Delta y^-)\right)\right],\\
    \end{cases}
\end{aligned}
\label{eq:entropy-spinning-defect}
\end{equation}
where a minimization is needed over candidate extremal surfaces in the bulk spacetime.
In any case, we see that each candidate, and thus the true RT surface, has length diverging with $\ell$.\footnote{From the discussion below~\eqref{eq:M-J-from-rpm}, $b_+$ also scales with $\ell$ in the flat limit, so most of the $\ell$ dependence inside the logarithms cancels out.} We again conclude
that the RT surface area is not well-defined in  the flat limit.

A natural objection to this conclusion is that properly understanding holographic entanglement entropy in the flat limit may require a more subtle manipulation than a simple $\ell\to\infty$ limit of the RT surface. In particular, the main obstruction to the usual RT story working nicely for subregions of null infinity is that geodesics terminating at null infinity are necessarily null. One can of course introduce a screen within Minkowski space on which to anchor spacelike geodesics and trade the issues above for ones of making sense of the theory on such a screen.
Since the  
original flat-space entanglement entropy investigations 
in \cite{Bagchi:2014iea, Jiang:2017ecm} aim to relate bulk geometric to Carrollian CFT computations, we defer discussions of screens at finite distance.
Also, we comment that regulating the RT entropy through a subtraction scheme (e.g.~subtracting the local endpoint divergence) will still leave a contribution proportional to $\ell$, and hence cannot cure the divergence.
We now turn to the boundary incarnation of this  flat limit issue.

\subsection{Boundary perspective}
\label{sec:bdy-persp}

In the previous subsection, we saw that RT surfaces in AdS$_3$ become ill-defined in the bulk $\ell\to\infty$ limit.
In this section, we discuss the boundary interpretation of this result. In particular, we would like to understand how the divergence in~\eqref{bmsee2holo} arises from the perspective of boundary twist operator correlation functions.

To do so, we first briefly summarize the twist-operator technique of Cardy and Calabrese \cite{Calabrese:2004eu,Calabrese:2009qy} in ordinary CFT$_2$'s. 
To use this technique, one begins with the following formula
\begin{equation}\label{replicalim}
S(A)
=
-\lim_{n\to 1}
\partial_n
\log
\tr(\rho_A^{n})\,,
\end{equation}
for the entanglement entropy $S(A)$ in terms of the vacuum density matrix $\rho_A$ associated with the region $A$.
Then, the relevant result of Cardy and Calabrese is that the object $\tr(\rho_A^{n})$ --- and hence the full entropy $S(A)$ --- is uniquely determined by two-point functions of certain conformal primary operators, 
as  summarized by the following formula:
\begin{equation}
\tr \left(\rho_A^n\right)
=
\left\langle
\sigma_n(\partial A_1)
\tilde{\sigma}_{n}(\partial A_2)
\right\rangle\,.
\label{cc}
\end{equation}
Here, 
$\partial A_i$ are the two endpoints of the interval $A$,
and $\sigma_n$ and $\tilde{\sigma}_{n}$ are twist and anti-twist operators,
namely conformal primary operators which both have the following scaling dimensions: 
\begin{equation}
    h^{(n)}=\frac{c}{24}
\left(
n-\frac{1}{n}
\right)
\qquad 
\bar h^{(n)}
=
\frac{\bar{c}}{24}
\left(
n-\frac{1}{n}
\right)\,.
\end{equation}
Given these weights, one readily obtains the entanglement entropy formula~\eqref{SCFT} from equations~\eqref{replicalim} and \eqref{cc}.

The question is then how to port this over to the Carrollian limit. 
As we will encounter here and in the next section, there are subtleties in how to do this. 
In the work of~\cite{Bagchi:2014iea}, for example, the authors arrive at a proposed Carrollian BMS field theory (BMSFT) entanglement entropy through a nontrivial trick involving an intermediate Galilean contraction. 
In the end, they obtain
BMS twist operators characterized by 
\begin{equation}
h_L^{(n)}
=
\frac{c_L}{24}
\left(
n-\frac{1}{n}
\right)\qquad
 h_M^{(n)}
  =\frac{c_M}{24}
\left(
n-\frac{1}{n}
\right)\,,
\label{bmsweight}
\end{equation}
and the entanglement entropy
\begin{equation}
S(A) = \frac{c_L}{6} \ln \left(\frac{2}{\varepsilon} \sin \frac{\Delta \phi}{2}\right)+\frac{c_M}{12}{\Delta u} \cot \frac{\Delta \phi}{2}
\,.
\label{bmsee}
\end{equation}
They assign the values of $c_L$ and $c_M$ as in \eqref{cstate} to comply with asymptotic symmetry analyses.

While we will keep the entropy formula \eqref{bmsee},\footnote{Indeed, see the independent check of the $c_M=0$ formula in \cite{Hao:2025hfa}.} we will argue that the  choice \eqref{cstate} for $c_L$ and $c_M$ is actually incorrect if we wish to reproduce the Carrollian limit of ordinary
CFT twist operator correlators.
In particular, we will argue below that the BMS subalgebra of Vir$\times$Vir  relevant to the question of CFT twist operator insertions at null infinity is not~\eqref{bms} but instead the following:
\begin{equation}
    \mathcal{L}'_n \equiv L_n + (-1)^n\bar{L}_n
    \qquad 
    \mathcal{M}'_n \equiv\frac{1}{\ell}\left(L_n - (-1)^n\bar{L}_n\right)\,,
    \label{gal}
\end{equation}
with central charges
\begin{equation}
    c_L' = c+\bar{c}
    \qquad
    c_M' =\frac{1}{\ell}(c-\bar{c})
    \label{bmscent2}
\end{equation}
differing from those in \eqref{cstate}.
Under this new contraction, Virasoro primary representations limit directly to the BMS primary representations used in \cite{Bagchi:2014iea}, and in particular, the CFT twist operators limit to BMS primaries with weights \eqref{bmsweight}.
This is then formally consistent with \eqref{bmsweight}-\eqref{bmsee}.

Crucially, however, with our new subalgebra, the holographic central charges read
\be\label{cl}
c_L'=\frac{3\ell}{G},~~c_M'=\frac{3}{\mu\ell G}\,,
\ee
instead of the central charge assignment made in \cite{Bagchi:2014iea}.
Due to the $\ell$ dependence in $c_L$, one can directly verify that~\eqref{bmsee} with central charges \eqref{bmscent2} formally reproduces the divergent RT result discussed in~\eqref{bmsee2holo}-\eqref{bmsee2holo_subleading}, providing the desired boundary interpretation of our previous bulk calculation.\footnote{Note, however, that in a non-holographic CFT, $c$ and $\bar{c}$ need not scale with $\ell$, and hence the associated Carrollian entropy need not diverge.}

It remains to reconcile this correct matching of the RT entropy with the mismatch between \eqref{bmscent2} and the central charges \eqref{cstate} expected for our AFS spacetime. 
This will require understanding precisely why the contraction \eqref{gal}, and not \eqref{bms}, is physically relevant for studying the Carrollian limit of local operators.
We now do so by studying how different choices of conformal frame in the parent CFT  can influence which contraction is required for describing the Carrollian limit.

\subsection{Central charge mismatch and a change of frame}
\label{sec:vectors}

It is useful to ground these discussions by looking at explicit vector field representations. 
Introducing the null coordinates~\eqref{eq:null}, we can write vector fields representing the Witt algebra as follows:
\be\label{lpm}
L_n=ie^{in y^+}\p_+,~~~\bar L_n=-ie^{-in y^-}\p_-.
\ee
The contraction~\eqref{bms} then gives the following BMS vector fields
\begin{equation}\label{vec1}
\begin{aligned}
{\cal L}_n & =i e^{i n \phi}\left(\partial_\phi+i n u \partial_u\right),~~
{\cal M}_n & =i e^{i n \phi} \partial_ u \,, 
\end{aligned}
\end{equation}
in the $\ell\to\infty$ limit. This is the standard In\"on\"u-Wigner contraction introduced in section~\ref{sec:basics}. 
As we will explain, it is the natural contraction for studying the symmetries of cylinder states in the Carrollian limit, 
and so we will refer to it as the {\it state-frame contraction}.
Indeed, we can already notice that none of these vector fields fix the origin $u=0$, $\phi=0$ (or any other finite point, for that matter)  and hence they are not immediately amenable to discussing symmetries acting on local operators.
For that, we argue that one must turn to the contraction \eqref{gal}.\footnote{
Note that $\bar{L}_n \to \alpha^n \bar{L}_n$ is an automorphism of the Virasoro algebra,
and hence by changing conventions for the definitions of the generators, we can turn \eqref{gal} into
\be \label{gal2}
{\cal L}'_n = L_n + \alpha^n \bar{L}_n, ~{\cal M}'_n = \frac{1}{\ell}\left(L_n - \alpha^n \bar{L}_n\right).
\ee
The case with $\alpha=1$ appears in~\cite{Bagchi:2014iea}, where it was interpreted as a Galilean contraction. 
This was due to the fact that exchanging time and space on the cylinder $y^-\mapsto -y^-$ formally flips $\bar L_n\mapsto -\bar L_{-n}$ according to~\eqref{lpm}. 
By contrast, we will here interpret \eqref{gal} with $\alpha=-1$ as implementing a Carrollian limit.
In appendix~\ref{app:more-bms}, we provide a more systematic treatment of the space of BMS$_3$ subalgebras.}

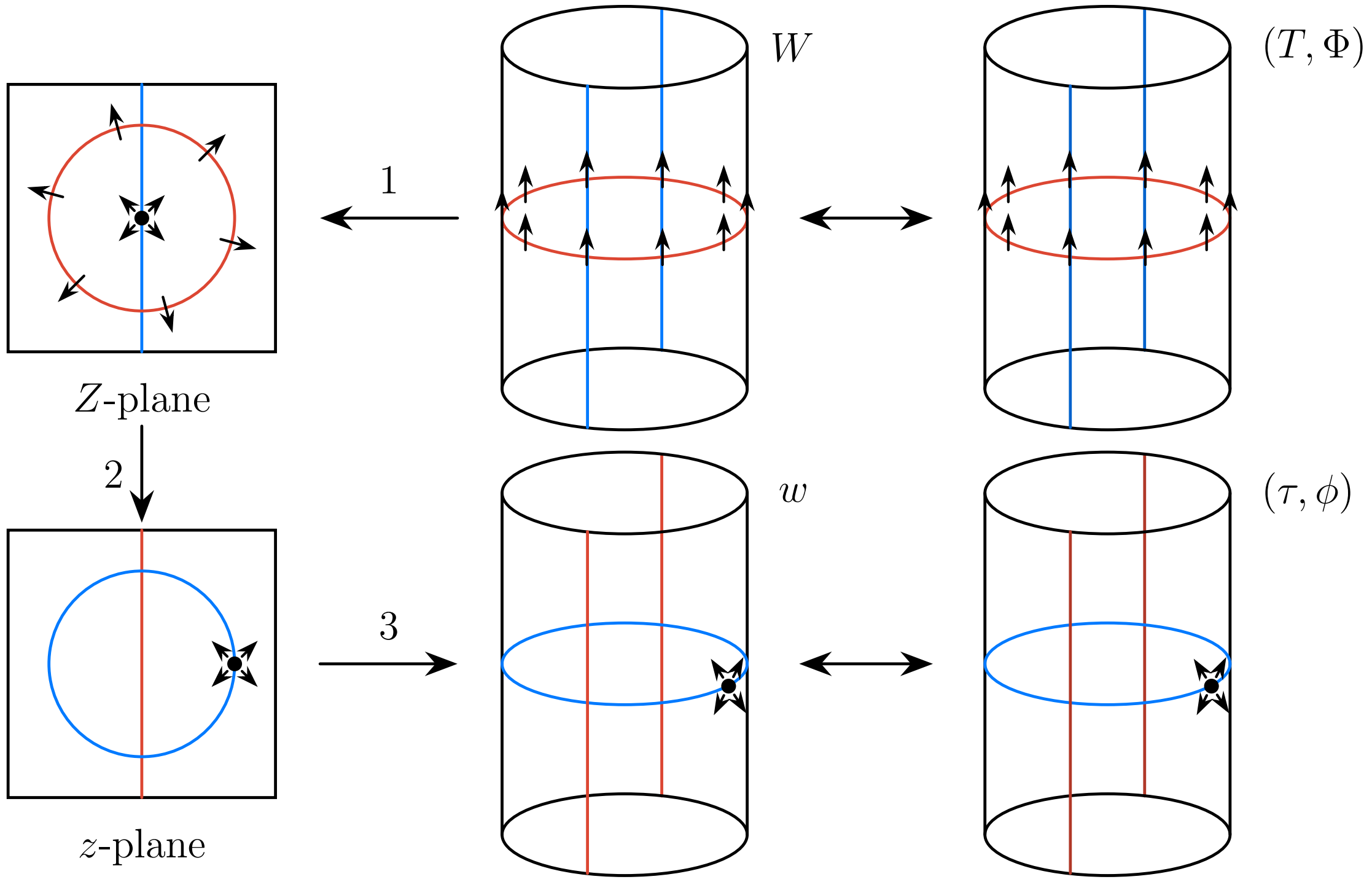
\begin{figure}[ht]
\centering
\hspace{1.5em}
\begin{tikzpicture}[scale=.8]

\coordinate (C10) at (0,0); 
\coordinate (C12) at ($(C10)+(\ColSep,0)$); 
\coordinate (C00) at ($(C10)+(0,-\RowSep)$); 
\coordinate (C02) at ($(C12)+(0,-\RowSep)$); 
\coordinate (C01) at ($(C10)+(-\ColSep,0)$); 
\coordinate (C-1) at ($(C00)+(-\ColSep,0)$); 


\DrawSquarePanel{(C01)}{myblue}{myred}{f}{t}
\node[labeltxt, below=5pt]
at ($(C01)+(0,-\SqSize/2)$)
{$Z\text{-plane}$};

\CuteArrowsAt{(C01)}
\CuteArrowsWIDE{(C01)}

\DrawCylinder{(C10)}{myblue}{myred}{f}
\node[right=.5pt,labeltxt]
at ($(C10)+(\CylR+0.1,\CylH/2)$)
{$W$};

\UpArrowsOnEllipse{(C10)}{\CylR}{\TopDepth}{10}{0}

\DrawCylinder{(C12)}{mydarkblue}{myred}{f}
\node[right=3pt,labeltxt]
at ($(C12)+(\CylR+0.1,\CylH/2)$)
{$(T,\Phi)$};

\UpArrowsOnEllipse{(C12)}{\CylR}{\TopDepth}{10}{0}

\draw[guide2]
($(C10)+(\CylR+0.7,0)$)
--
($(C12)+(-\CylR-0.7,0)$);

\draw[guide]
($(C10)+(-\CylR-0.6,0)$)
--
($(C01)+(\SqSize/2+0.6,0)$)
node[midway, above=3pt, labeltxt] {1};


\DrawSquarePanel{(C-1)}{myred}{myblue}{t}{f}
\node[labeltxt, below=5pt]
at ($(C-1)+(0,-\SqSize/2)$)
{$z\text{-plane}$};

\CuteArrowsAt{($(C-1)+(\CircR,0)$)}

\DrawCylinder{(C00)}{myred}{myblue}{t}

\CuteArrowsAtCYL{($(C00)+(\CylR-.25,-.3)$)}

\node[right=3pt,labeltxt]
at ($(C00)+(\CylR+0.1,\CylH/2)$)
{$w$};

\DrawCylinder{(C02)}{mydarkred}{myblue}{t}

\CuteArrowsAtCYL{($(C02)+(\CylR-.25,-.3)$)}

\node[right=3pt,labeltxt]
at ($(C02)+(\CylR+0.1,\CylH/2)$)
{$(\tau,\phi)$};

\draw[guide2]
($(C00)+(\CylR+0.7,0)$)
--
($(C02)+(-\CylR-0.7,0)$);

\draw[guide]
($(C-1)+(\SqSize/2+0.6,0)$)
--
($(C00)+(-\CylR-0.6,0)$)
node[midway, above=3pt, labeltxt] {3};

\draw[guide]
($(C01)+(0,-\SqSize/2-1)$)
--
($(C-1)+(0,\SqSize/2+0.1)$)
node[midway, left=1pt, labeltxt] {2};

\end{tikzpicture}

\caption{
The conformal transformation between the state frame and the local-operator frame.  
In Euclidean signature, it is the composition of a cylinder-to-plane map, a M\"obius transformation, and a plane-to-cylinder map. 
It sends the lower asymptotic end of the $W$ cylinder to the point $w=0$. 
The two Lorentzian cylinders are obtained by Wick rotating different directions (red vs.~blue) in the complexified geometry; they are therefore different real slices rather than two coordinate systems on conformally equivalent Lorentzian cylinders. 
}
\label{fig:stateop}
\end{figure}

We begin with an explanation of why \eqref{vec1}-\eqref{lpm}  are natural for studying the symmetries of cylinder states, before performing a change of conformal frame to study the symmetries of local operators in the Carrollian limit.

\paragraph{Symmetries in the state frame}
Consider a relativistic CFT before any Carrollian limits. States on the cylinder are characterized using the map from the cylinder to the plane.
With a Wick rotation
\be
w=iy^+ \qquad \bar{w}=-iy^-\,,
\ee
the conformal transformation from the cylinder to the plane reads
\be\label{ctop}
z=e^{w},~~\bz =e^{\bar{w}}\,,
\ee
and the vector fields in~\eqref{lpm} become the standard Virasoro vector fields on the plane:
\be 
L_n=-z^{n+1}\p_z,~~~\bar L_n=-\bz^{n+1}\p_\bz \,.
\label{lpmPlane}
\ee
These symmetries can be used to characterize a local primary operator placed at $z=\bar{z}=0$, 
the point which maps under \eqref{ctop} to the bottom asymptotic end of the $(w,\bar{w})$ cylinder.

Under this map, the highest weight condition is the same. 
Hence, by acting with primary operators $\mathcal{O}_{h,\bar{h}}$ at $z=\bz=0$,
we can prepare primary states $|h,\bar h\rangle$ on the cylinder, which obey
\be \label{primconditions}
L_0|h,\bar h\rangle=h|h,\bar h\rangle,~~\bar L_0|h,\bar h\rangle=\bar h|h,\bar h\rangle,~~~L_n|h,\bar h\rangle=\bar L_n|h,\bar h\rangle=0~~ \forall~ n>0\,.
\ee
Other states in the representation are formed by acting with the lowering operators $L_n$, $\bar{L}_n$, $n< 0$. 

Crucially, the Virasoro generators  appearing in the cylinder relations \eqref{primconditions} are explicitly associated with the symmetries described by the \eqref{vec1} vector fields.
To describe these symmetries in the Carrollian limit, one must identify linear combinations of those vector fields which remain finite in the limiting Carrollian time-band.
This finiteness requirement is highly constraining, and one finds that the state-frame contraction \eqref{bms} is essentially the unique option up to automorphisms. 

Hence, the state-frame contraction \eqref{bms} with its finite central charges \eqref{cstate} provides the natural way of taking the Carrollian limit in the conformal frame characterized by the vector fields \eqref{vec1}.
Under this contraction, usual Virasoro primaries $|h,\bar{h}\rangle$ limit to a somewhat complicated representation of the BMS algebra,
due to the mixing between positive and negative Virasoro modes in \eqref{bms}. 
More about the resulting {\it induced representations} can be found in the pedagogical reference \cite{Oblak:2016eij}, as well \cite{Campoleoni:2016vsh,Hao:2025btl}. For our purposes, we  need only point out that these states are not the {\it BMS primary representations} appearing in~\cite{Bagchi:2014iea}, which we define shortly.

\paragraph{Mapping to the operator frame}
We now carry out the change of frame.
From our exponential map \eqref{ctop}, we see that state preparation corresponds to inserting operators at $z=\bz=0$, whereas placing a primary operator at the 
$u=0, ~\phi\in(0,2\pi)$ time-slice on the cylinder (i.e. $y^+=y^-=\phi$) corresponds to inserting an operator on the unit circle in the $z$-plane. 
To relate these cases, we introduce the M\"obius transformation
\begin{equation}
    Z=\frac{z-1}{z+1}\,.
    \label{mobius}
\end{equation}
This relates $Z=0$ to $z=1$, as desired, and also relates the unit circle in the $Z$-plane to the imaginary axis of the $z$-plane. 

This M\"obius transformation can be enacted by conjugating with
\be\label{Wmap}
 W= e^{\frac{\pi}{4}(L_1+L_{-1})}
    e^{\frac{\pi}{4}(\bar{L}_1+ \bar{L}_{-1})}\, .
\ee
Upon composing $W$ with the exponential map \eqref{ctop}, which we leave implicit, we can act with the change of frame $W$ on our cylinder Virasoro generators \eqref{lpm}. Namely we define $\ell_n:= W L_n W^{-1}$ and $\bar \ell_n:= W \bar L_n W^{-1}$. Then, 
using the explicit coordinate transformation, we find a new set of Witt vector fields on the cylinder:
\begin{equation}
\begin{aligned}
    \ell_n &= -  \sin y^+ \left(i \tan (y^+/2)\right)^n \,\p_{+}
    \qquad
    \bar{\ell}_n = -   \sin y^- \left(- i \tan (y^-/2)\right)^n \,\p_{-}\,.
    \label{sinhtanh}
\end{aligned}
\end{equation}
In a Taylor expansion near the origin, these resemble the more standard Virasoro vector fields \eqref{lpmPlane}, and in particular, they fix the origin for $n\ge0$. 

Again, to describe these symmetries in
the Carrollian limit, one must identify linear combinations which describe finite vector fields in the limiting Carrollian time-band. This finiteness requirement is again highly constraining, 
and leads to the combinations from \eqref{gal}, namely
\be\label{opframe}
  {\cal L}_n' = \ell_n + (-1)^n \bar{\ell}_{n},~~
        {\cal M}_n' = \frac{1}{\ell}\left(\ell_n - (-1)^n \bar{\ell}_{n} \right)\,.
\ee
In particular, the associated  Carrollian vector fields read 
\begin{equation}\label{vec2}
    {\cal L}_n' = -\left( i \tan \frac{\phi}{2}\right)^n\left( \sin\phi\, \p_\phi +  u (n+\cos\phi) \p_u \right)
    \qquad
    {\cal M}_n' = -\left(i \tan \frac{\phi}{2}\right)^n \sin\phi \,\p_u \, ,
\end{equation}
and they fix the origin $u=\phi=0$ for $n\ge 0$. 
We see that, as claimed, the contraction \eqref{gal} is relevant for studying the Carrollian limit of local operators, and we accordingly refer to it as the {\it operator-frame contraction}.

Since M\"obius transformations do not change the primary conditions,
and since \eqref{opframe} does not mix modes with positive and negative labels, the conjugated Virasoro primaries
$\mathcal{O}_{h_L, h_M}'(0,0):=W\mathcal{O}_{h, \bar{h}}W^{-1}$
will now be BMS primaries, defined via
\begin{equation}
\begin{aligned}\label{bmsprim2}
& [{\cal L}'_0, \mathcal{O}'_{h_L, h_M}]=h_L \mathcal{O}_{h_L, h_M}',~~ 
[{\cal M}'_0, \mathcal{O}'_{h_L, h_M}]=h_M \mathcal{O}_{h_L, h_M}' \\
& [{\cal L}'_n,\mathcal{O}'_{h_L, h_M}]=[{\cal M}'_n,\mathcal{O}'_{h_L, h_M}]=0 \text { for } n>0 \,,
\end{aligned}
\end{equation}
and pedagogically reviewed in \cite{Riegler:2016hah}.
Here, the inherited weights are
\be
h_L= h+\bar h,~~ h_M=\frac{1}{\ell}(h-\bar h)\,,
\ee
and the central charges of this BMS subalgebra, assuming a holographic parent CFT, are the ones in~\eqref{cl}.
This explains, in particular, that CFT twist operators in holography limit to BMS primaries with the ill-defined weights specified in \eqref{bmsweight} with \eqref{cl}, 
which was the key step in explaining the divergence of the RT formula in the previous section.

\vspace{1em}\noindent
Stepping back, we observe that the reason why we landed on a different BMS subalgebra in each frame is that we conjugated with a symmetry $W$ which fails to limit to a finite BMS transformation. That is, we can use~\eqref{bms} to recast the CFT state-operator map~\eqref{Wmap} in terms of BMS symmetries:
\be
W=e^{\frac{\pi}{4} \ell ({\cal M}_1+{\cal M}_{-1})},
\ee
and we note that the explicit $\ell$-dependence renders $W$ ill-defined in the Carrollian limit.\footnote{There nevertheless exists a formal \cite{Chen:2024voz} Carrollian state-operator map $\mathcal{W}$, given by
$e^{i\Phi}=i\tan(\phi/2)$ and $iU=u\csc\phi$ 
\cite{Bagchi:2015nca,Hao:2021urq}.
Its lift to the parent CFT  fails to preserve the Euclidean reality conditions. Relative to the ordinary Euclidean state-operator map, it involves the reflection $y^-\mapsto\pi-y^-$. The corresponding formal mode redefinition $\widetilde{\bar L}_n=(-1)^{n+1}\bar L_{-n}$ yields a Virasoro algebra with central term proportional to $-\bar c$. This provides a complementary perspective on the central charge mismatch, at the price of leaving the Euclidean section of the parent CFT.}
The two central-charge assignments \eqref{cstate} vs.~\eqref{bmscent2} should therefore not be regarded as an ambiguity within the same limiting theory; instead, one obtains different Carrollian theories  depending on the choice of conformal frame in the parent CFT.
Put differently, if we conjugate a BMS representation with a global BMS transformation, we will not change the  BMS primary condition. Similarly, if we conjugate a Virasoro primary with a global conformal transformation, we will not change the CFT primary condition. 
However, conjugating by Virasoro transformations can cause a different combination of Vir $\times$ Vir generators to yield BMS$_3$ in the $\ell\to\infty$ limit, changing the emergent BMS representations and central charges.

To summarize, there exists a CFT change of frame taking states to operators at $u=\phi=0$.
These operators transform as BMS primaries in the flat limit, albeit under a subalgebra with divergent holographic central charges differing from \eqref{cstate}.
This produces the alternate contraction~\eqref{gal} and central charges~\eqref{cl} needed to make~\eqref{bmsee} match the divergent RT result.
Meanwhile, the finite swing formula is obtained by swapping out these BMS central charges for those associated with a standard flat space asymptotic symmetry analysis.
In the discussion section, we speculate about possible mechanisms for how one could arrive at a local operator algebra in the Carrollian limit governed by the physical flat-space central charges.\footnote{Of course, one could employ the state-frame contraction, followed by the BMS state-operator map $\mathcal{W}$ introduced in the previous footnote, to engineer local operators transforming under a BMS algebra with finite central charges.
However, this route is rather unappealing: as mentioned above, $\mathcal{W}$ violates the Euclidean reality conditions, and moreover, it appears one would obtain induced representations instead of highest weight representations at the end of this procedure, and this is a setting where twist operator computations are understudied.}

\section{Swing surfaces in AdS/CFT}
\label{sec:swing}

In the previous section, we demonstrated that RT surfaces are ill-defined in the flat-space limit.
We also showed that, with some care regarding the change of frame, one recovers this divergent result through the Carrollian limit of boundary twist operators. 
Meanwhile, the Carrollian ``entanglement entropy'' of \eqref{bmsee} is well-defined in the flat limit, though not being manifestly an entanglement entropy.
It reads 
\begin{equation}
    S_{\textrm{swing}}(A) = \frac{1}{4 G}\Delta u \cot \frac{\Delta \phi}{2}
    +
    \frac{1}{2 \mu G} \ln \left(\frac{2}{\varepsilon} \sin \frac{\Delta \phi}{2}\right)\,,
    \label{eq:swing-len}
\end{equation}
and it has a conjectured bulk dual via the swing surface proposal. 

In the remainder of this work, 
our goal will be to understand if and how the swing proposal can be situated in AdS/CFT.
To initiate our study of swing surfaces, in this section, we introduce ordinary swing surfaces in flat-space holography.
Then, we describe a swing-type surface in AdS which limits to its flat-space counterpart in the $\ell\to\infty$ limit.
We provide a consistency check on our uplifted swing surface, based on the field theory  result in section \ref{sec:bdy-persp}.
Our explicit bulk construction will be restricted to Einstein gravity ($\mu\to\infty$) where only the first term in
 \eqref{eq:swing-len} remains.  
However, the TMG term can be recovered from our construction using similar normal-frame twist techniques to that in section 5 of~\cite{Hijano:2017eii}. 

\subsection{Bulk construction}
\label{sec:bulk-persp-2}

Let us introduce the swing proposal in flat-space holography.
One begins with three-dimensional Minkowski spacetime and its future asymptotic boundary $\mathcal{I}^+$.
As discussed below \eqref{carbdymetric}, 
$\mathcal{I}^+$ can be thought of as a Carrollian cylinder with a null direction $u$ and a periodic spatial dimension, $\phi$.
The basic assumption underlying the swing proposal is that the holographic dual of three-dimensional flat-space gravity is a quantum theory living on this two-dimensional Carrollian manifold.

The starting point of Wei Song and collaborators~\cite{Jiang:2017ecm} was then to look for a bulk quantity equal to the purported BMSFT entanglement entropy for an interval of size $(\Delta u, \Delta \phi)$ on $\mathcal{I}^+$ proposed by Bagchi et al.~\cite{Bagchi:2014iea}; namely,~\eqref{eq:swing-len}. Note that throughout this section, we restrict to the case of Einstein gravity in the bulk, so we have $c_M=3/G$ and $c_L=0$ in \eqref{eq:swing-len}.
The swing proposal is then the statement that
\begin{equation}
    S_{\textrm{swing}}(A) = \frac{\textrm{Length}(\gamma_\textrm{swing})}{4G}\,,
\end{equation}
where $\gamma_\textrm{swing}$ is an extremal surface in the bulk Minkowski spacetime anchored to radial null rays emanating from $\p A\equiv\{\partial A_1, \p A_2\}$. 
Explicitly, one begins by shooting 
one radial past-directed null geodesic from each endpoint of the interval $A$ on $\mathcal{I}^+$, and then one extremizes over the endpoints tethered to these fixed null ``ropes.'' 
The length of the swing surface was originally computed in \cite{Jiang:2017ecm} and found to match \eqref{eq:swing-len}.
The fact that the ropes are radial introduces a strange coordinate dependence that we aim to clarify with our AdS constructions.

Our main goal in this section is to provide a swing-like surface in AdS$_3$ which limits to its flat-space counterpart in the $\ell\to\infty$ limit.
Our goal in later sections will be to explore whether this surface has a meaningful origin in the context of AdS/CFT.

The prescription from section~\ref{sec:basics} already tells us where to lift the endpoints of the interval to on the Lorentzian cylinder, namely to $\tau=\frac{\pi}{2}+u_i/\ell$, and $\phi=\phi_i$ so that $(\Delta\tau,\Delta\phi)=(\Delta u/\ell, \Delta\phi)$.
Then, mimicking the swing proposal, 
we define two null ropes by shooting one past-directed null geodesic into the bulk from $\partial A_1$ and one past-directed null geodesic into the bulk from $\partial A_2$.
These should be radial null rays, with respect to the coordinates in \eqref{uads}.
Finally, we define our swing surface $\gamma_{\textrm{swing}}$ by extremizing over surfaces with endpoints along the null ropes.  
From the formalism in section \ref{sec:basics}, it is clear that the uplifted swing surface limits to the original swing surface in the $\ell\to\infty$ limit.

Towards a CFT understanding of this object, 
let us obtain explicit formulas describing 
the swing surface at finite $\ell$.
To do so, 
we first note that in the flat-space context, and in the vacuum setting,
there is a unique candidate for
$\gamma_{\textrm{swing}}$,
and it is contained in the intersection of the past-directed lightsheets emanating from $\partial A_1$ and $\partial A_2$ \cite{Jiang:2017ecm}.
At finite $\ell$ in global AdS, it is also true that
$\gamma_{\textrm{swing}}$ is contained in the intersection of the past-directed lightsheets emanating from $\partial A_1$ and $\partial A_2$, and that $\gamma_{\textrm{swing}}$ furnishes a unique local extremum.
This makes it easy to obtain an explicit formula for the swing surface.
Specifically, in the geometry \eqref{ads}, the past lightsheet emanating from a boundary point $(\tau_*, \phi_*)$ 
is given by 
\begin{equation}
\cos\left({\tau-\tau_*}\right) 
= \frac{1}{\sqrt{\ell^2/r^2+1}}\cos(\phi-\phi_*)\,.
\end{equation} 
The intersection of the two lightsheets emanating from $\partial A_i=(\frac{\pi}{2}\pm \frac{\Delta u}{2\ell},\pm \frac{\Delta \phi}{2})$ then reads
\begin{equation}
\begin{aligned}
\tau(\phi)
        &=
    \tan^{-1}\left[\frac{\tan\frac{\pi-\Delta\phi}{2}}{\tan\frac{\pi-\Delta u/\ell}{2}}\tan(\phi-\frac{\pi}{2})\right]\\
    r(\phi)^2 &= \frac{\ell^2 \sin^2(\pi-\frac{\Delta u}{\ell})}{\left(\cos(\pi-\Delta \phi)-\cos(\pi-\frac{\Delta u}{\ell})\right)\left(\cos(2(\phi-\frac{\pi}{2}))-\cos(\pi-\frac{\Delta u}{\ell})\right)}\,
    \label{int}\,.
\end{aligned}
\end{equation}
The swing surface for $A=(\Delta \tau,\Delta \phi)$ is the portion of this geodesic which extends between the null-rope cutoff values, schematically denoted
$\phi_1= -\Delta \phi/2$, 
and
$\phi_2= \Delta \phi/2$.
Note that by comparing \eqref{int} with \eqref{geod-og}, we see that the curve in \eqref{int} simply describes a spacelike geodesic in AdS$_3$ centered at $(\tau_0,\phi_0)=(0,\pi/2)$, with boundary opening angle $\Delta \Phi $ and time extent $\Delta T$ given by
\begin{equation}
\Delta \Phi = \pi-\frac{\Delta u}{\ell}
\qquad
    \Delta T = \pi-\Delta \phi
    \,.
    \label{eq:swing-from-rt}
\end{equation}
To compute the length of the swing surface, 
we simply compute the length of the portion of this geodesic between the two angles $\pm \Delta\phi/2$  set by the radial cutoffs.
This gives
\begin{equation}
    \text{Length}(\gamma_{\textrm{swing}}) =
   2\ell \,\text{tanh}^{-1}\left[\frac{\tan \frac{\Delta u}{2\ell}}{\tan \frac{\Delta \phi}{2}}\right]
   \label{eq:swing-len-ads}
\end{equation}
as the length of the swing surface in AdS$_3$; 
here and below we suppress the absolute values required to make sure lengths are positive.
Note  that  after taking $\ell\to\infty$, we recover the length associated with the swing surface in Einstein gravity, corresponding to the first term of \eqref{eq:swing-len}.
However, part of the point of this discussion is that 
nothing stops us from keeping $\ell$ finite, i.e. taking $\tau_i$ outside of the band near $\tau\sim \frac{\pi}{2}$, and treating the associated swing bench as a genuine observable in AdS/CFT.

To close this section, we comment that the swing proposal in flat-space holography admits various refinements. 
For example, promising calculations have been made outside the vacuum setting \cite{Apolo:2020bld}.
One subtlety is that when we go beyond the vacuum setting, one nominally needs to minimize over multiple candidate swings; however, 
certain modular flow arguments~\cite{Jiang:2017ecm} 
capture the outermost swing candidate, even when it differs from the minimal-length candidate.

Putting this subtlety aside, 
the calculations in \cite{Apolo:2020bld} invite us to briefly discuss swing surfaces in backgrounds which differ from global AdS$_3$, and compare them to the answers which have been obtained in the flat limit. For this purpose, the simplest examples one can consider are spinning defect geometries described in \eqref{eq:defect-metric-bpm}.
In appendix \ref{app:spsm_check}, we study swing surfaces for the geometries in \eqref{eq:defect-metric-bpm}.
In general, there are two relevant swing candidates.
The first candidate has the following length:\footnote{Note that in \eqref{eq:entropy-diff-spinning-defect}, the first term 
is just a rewriting of the vacuum swing length \eqref{eq:swing-len-ads} 
under the change of coordinates which relates vacuum AdS$_3$ to the spinning defect geometry (e.g. see \cite{Caminiti:2025gyv,Li:2024rma}).
Meanwhile, the second term  corrects for the fact that this 
change of coordinates does not preserve the radial property for null rays.}
\begin{equation}
\begin{aligned}
   \textrm{Length}(\gamma_{\textrm{swing}})=
    2\ell\text{tanh}^{-1}\frac{\tan\frac{b_+\Delta \tau + b_- \Delta \phi}{2 \ell}}{\tan\frac{b_- \Delta \tau + b_+ \Delta \phi}{2 \ell}}
    +
   \ell \log\frac{b_+-b_-}{b_++b_-}
    \,,
\end{aligned}
\label{eq:entropy-diff-spinning-defect}
\end{equation}
while the second candidate simply has $\Delta\phi$ replaced by $2\pi-\Delta\phi$ and $\Delta \tau\to-\Delta\tau$.
Now, the flat-space analog of this formula was computed in \cite[appendix D]{Apolo:2020bld},
and it is straightforward to check that it indeed matches the $\ell\to\infty$ limit of \eqref{eq:entropy-diff-spinning-defect} when the parameters $M$ and $J$ are held fixed.
Again, however, the main point is that nothing stops us from keeping $\ell$ finite and studying swings in AdS/CFT.

\subsection{Boundary construction}
\label{sec:bdy-persp-2}

So far, we have defined the swing surface in flat space, as well as its natural uplift to AdS$_3$.
In particular, we derived the length of the uplifted swing surface in the vacuum state in \eqref{eq:swing-len-ads}.
In this subsection, we will examine the uplift of its boundary counterpart.

As we encountered at the end of section~\ref{sec:RTflat}, the difference between the usual entanglement entropy and the BMS quantity proposed by~\cite{Bagchi:2014iea} boils down to the particular assignment of values to the BMS central charges $c_M$ and $c_L$. 
In particular, the only difference between the two answers is a sign flip on $\bar{c}$, 
inviting us to compare AdS$_3$ swing lengths to the following  CFT quantity:
\begin{equation}
    S_{\textrm{swing}}(A) = S^+(A)-S^-(A)\,
    \label{spsm}
\end{equation}
where we induced a sign flip relative to the true entanglement entropy
\begin{equation}
    S(A) = S^+(A)+S^-(A)\, .
    \label{see}
\end{equation}
See also \cite{Riegler:2014bia} for a similar formula in the context of thermal entropy in flat holography.

Indeed, for a single interval in the AdS vacuum, we have
\begin{equation}
    \begin{cases}
        S^+(A)=\frac{c}{6} \log \left(\frac{2}{\varepsilon} \sin\frac{\Delta y^+}{2}\right)
        \\
        S^-(A)=\frac{\bar{c}}{6} \log \left(\frac{2}{\varepsilon} \sin\frac{\Delta y^-}{2}\right)
    \end{cases}
    \label{SPSM}
\end{equation}
and with the sign-flipped entropy we reproduce
\begin{equation}
    S_{\textrm{swing}}(A)
    =
   \frac{c}{6} \log \frac{\sin\frac{\Delta \phi+\Delta u/\ell}{2}}{\sin\frac{\Delta \phi-\Delta u/\ell}{2}}
   =
   \frac{c}{3}\tanh^{-1}\left[\frac{\tan\frac{\Delta u}{2\ell}}{\tan\frac{\Delta \phi}{2}}\right]
   \,,
\label{SCFT2}
\end{equation}
which is precisely $1/4G$ times the length \eqref{eq:swing-len-ads} computed above.
We emphasize that it was not obvious that these quantities had to match at finite $\ell$, and that this suggests a deeper meaning behind the relative sign flip between left- and right-moving sectors.

Nevertheless, one immediately finds subtleties with the swing proposal when written in this form. 
First, the length of the swing bench is positive by definition, whereas the sign-flipped quantity~\eqref{spsm} has no reason to be positive. 
Indeed, this negativity was already an issue for interpreting the signed object from~\cite{Bagchi:2014iea} as an entanglement entropy. Even restoring an absolute value, one should be somewhat wary of being able to isolate the left- and right-moving terms in a general state. 
For a single interval in the vacuum state, we indeed have such a factorization. 
If we look at the defect geometries, however, typically the minimization over candidates discussed below \eqref{eq:entropy-spinning-defect} obstructs a factorization of the entropy into a left- and right-moving piece as in \eqref{SCFT2},
as the minimization formula may depend nontrivially on both $\Delta y^+$ and $\Delta y^-$.

One can ask whether the same formula \eqref{spsm} at least holds in other states which do obey left-right factorization. Indeed, looking at \eqref{eq:entropy-spinning-defect} we see that when either $b_+ + b_-=\ell$ or $b_+-b_-=\ell$ , or equivalently
\begin{equation}
    M - M_0 = |J|/\ell\,, \qquad M_0 \equiv -1/(8G)\,.
    \label{eq:extremally-excited}
\end{equation}
The minimization will only affect one term, and hence we have a factorized result, e.g. for the $b_+ + b_-=\ell$ case
\begin{equation}
\label{facto}
    S^+(A) = \frac{c}{6}\log \left[\frac{2}{\varepsilon}\sin\left(\frac{\Delta y^+}{2}\right)\right]
    \qquad 
    S^-(A)=
    \text{min}
    \begin{cases}
        \frac{c}{6}\log \left[ \frac{2\ell}{\varepsilon(b_+-b_-)}\sin\left(\frac{b_+-b_-}{2 \ell}\Delta y^-\right)\right]\\
        \frac{c}{6}\log \left[ \frac{2\ell}{\varepsilon(b_+-b_-)}\sin\left(\frac{b_+-b_-}{2 \ell}(2\pi-\Delta y^-)\right)\right]\,.
    \end{cases}
\end{equation}
Equation~\eqref{eq:extremally-excited} says that the CFT state has been excited above the vacuum by an operator whose energy is equal to its spin. Namely such an operator has $h\ne 0$ yet $\bar{h}=0$ (or vice versa), 
and so it is not surprising that the resulting geometry indicates no entanglement  between left- and right-moving modes.
The question is then whether or not  $S^+(A)-S^-(A)$  matches the swing bench. 
Using the results discussed near \eqref{eq:entropy-diff-spinning-defect}, one can indeed show that each candidate in \eqref{facto} matches the length of a candidate swing bench.\footnote{It is readily verified that this candidate-by-candidate result also holds for general $b_+$ and $b_-$.}
However, it is puzzling to notice that taking the minimum of the absolute value  $|S^+-S^-|$ (to isolate one swing bench) need not be the same as taking the minimum in~\eqref{facto} (coming from the right-moving entropy).  

With these subtleties in mind, we may now turn to a more thorough investigation of the information-theoretic interpretation of~\eqref{spsm}. 

\section{Swing interpretation: take I}
\label{sec:lift}

In this section, we find that the swing surface in vacuum AdS$_3$ is dual to a boundary mutual information. 
While we focus on the vacuum case in section~\ref{sec:swingMI}, we later summarize, in section \ref{sec:swingGen},  the key conditions required for this identification to generalize.
Throughout this section, we will restrict to the case of Einstein gravity in the bulk and work at leading semiclassical order.

\subsection{Swing surface as a vacuum mutual information}
\label{sec:swingMI}

To show that the swing surface in vacuum AdS$_3$ is dual to a boundary mutual information,
we take~\eqref{spsm} as a starting point:
\begin{equation}
\frac{\textrm{Length}(\gamma_\textrm{swing})}{4G}=| S_{\textrm{swing}}(A)| = |S^+(A)-S^-(A)|\,.
\end{equation}
Here, we have taken the absolute value to guarantee a positive quantity on the right-hand side; in particular, both the bulk swing bench length and boundary mutual information will be nonnegative.

\begin{figure}[t!]
    \centering
        \centering
        \includegraphics[width=.7\textwidth,trim={0cm 0cm 0cm 0cm}, clip=true]{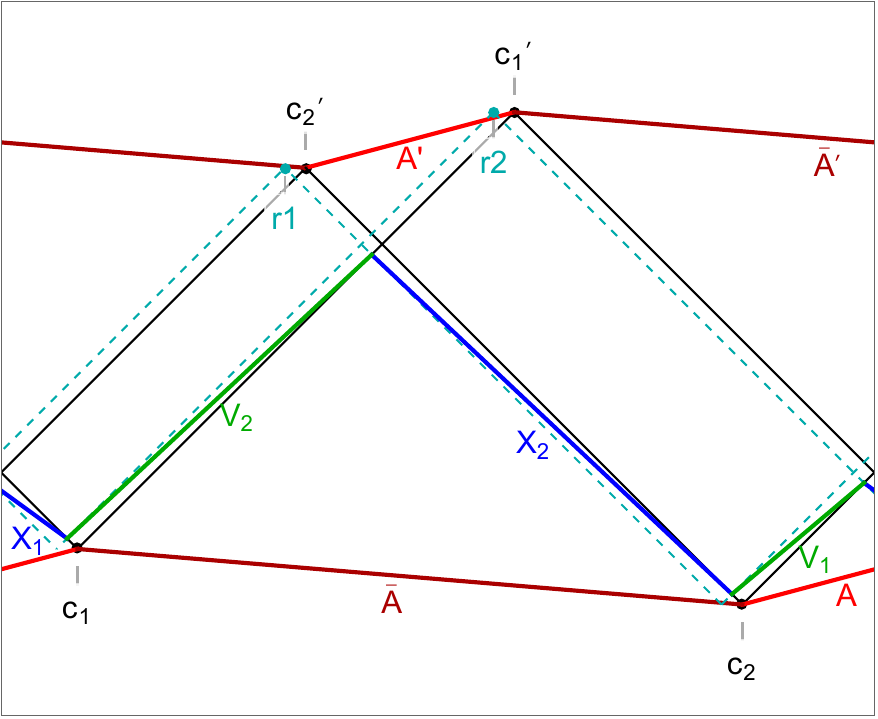}
    \caption{Regions on the Lorentzian cylinder used for identifying the swing length for the interval $A$ as a mutual information.
    All black lines lie at 45 degrees and correspond to null rays.} 
    \label{fig:regions2}
\end{figure}

In order to relate the swing length to a mutual information, 
the key step is to note that the left- and right-moving vacuum entropies $S^+(A)$ and $S^-(A)$ can be written in terms of ordinary vacuum entropies of nearly null intervals.
That is, \eqref{SCFT} implies
\begin{equation}
    S^+(A)=S(V_1)\,,~~~S^-(A)=S(X_1)\,,
    \label{fac}
\end{equation}
where we define the nearly null intervals $V_1$ and $X_1$
as follows:
\be\label{eq:VX1}
V_1: ~~(\Delta y^+-\varepsilon,\,\varepsilon)\,,~~ X_1:~~(\varepsilon, \,\Delta y^--\varepsilon)\,,
\ee
and we have chosen 
$\varepsilon$ to match the UV cutoff in \eqref{SCFT} so as to satisfy \eqref{fac}.\footnote{Identifying the null width with the UV cutoff is a convenient choice. However, in principle one may instead introduce an independent null-width parameter $\eta$, remove the UV cutoff at fixed $\eta$, and only then take $\eta\to0$. In Einstein gravity, the UV-cutoff dependence cancels in the entropy differences and mutual information considered here.
The important data is therefore not the cutoff itself but instead how our intervals approach null intervals.}
The intervals $V_1$, $X_1$ are shown in figure \ref{fig:regions2}.
We see that the surface $V_1\cup X_1$ is homologous to $A$ in the strict $\varepsilon\to 0$ limit.

At this stage, we have:
\begin{equation}
    S_{\textrm{swing}}(A) = S(V_1)-S(X_1)\,.
\end{equation}
To upgrade this difference of entropies to a mutual information, we introduce a second set of nearly null intervals $V_2$ and $X_2$ as follows:
\be\label{eq:VX2}
V_2: ~~(2\pi-\Delta y^+-\varepsilon,\,\varepsilon)\,,~~ X_2:~~(\varepsilon, \,2\pi-\Delta y^--\varepsilon)\,,
\ee
associated with (i.e. their union is parametrically homologous to) the complementary interval $\bar{A}$.

The fact that the CFT vacuum is a pure state gives us $S(A) = S(\bar{A})$. Meanwhile, from the fact that $S(A)$ factorizes as in \eqref{see},
we can straightforwardly derive the identities $S(V_1)=S(V_2)$ and $S(X_1)=S(X_2)$.
These identities allow us to rewrite $S_{\textrm{swing}}(A)$ as follows:
\begin{equation}\label{swingexp}
    S_{\textrm{swing}}(A) = \frac{1}{2}\left(S(V_1)+S(V_2)-S(X_1)-S(X_2)\right)\,.
\end{equation}
It turns out that this quantity is proportional to a mutual information.
To see this, we write out the definition
\begin{equation}
        I(V_1:V_2) 
    =
    S(V_1) + S(V_2) - S(V_1\cup V_2)\,.
    \label{mi}
\end{equation}
If we have $S(V_1)>S(X_1)$, then
$S(V_1)+S(V_2) \ge S(X_1)+S(X_2)$.
Consulting figure \ref{fig:regions2},
we see this is precisely the statement that $V_1\cup V_2$ has a connected entanglement wedge,
or in other words, $S(V_1\cup V_2)= S(X_1)+S(X_2)$ and hence we have~\eqref{mi} equal to twice~\eqref{swingexp}.
In the opposite case $S(V_1)<S(X_1)$, the $X_i$ and $V_i$ simply change roles.
Consequently, we obtain
\begin{equation}\label{eq:toshow}
   | S_{\textrm{swing}}(A)| = \frac{1}{2}\max\{I(V_1:V_2),I(X_1:X_2)\}\,.
\end{equation}
For either sign of $S_{\textrm{swing}}(A)$, one of the two cases vanishes,
so that the swing surface area is always related to a boundary mutual information.
This proves our claim that bulk swing surfaces are dual to boundary mutual informations in the vacuum state.

In the following subsection, we will discuss a simple bulk interpretation of this fact.
For this, it will be useful to instead view our interval $A$ as sitting at the future null infinity time band as $A'$ in figure \ref{fig:regions2}.  
Since $A'$ and $A$ are simply related by the translations $\phi\mapsto\phi+\pi$  and $\tau\mapsto \tau+\pi$, we have $S_{\textrm{swing}}(A)= S_{\textrm{swing}}(A') $. The discussions from this section carry through but we can now more readily relate them to a bulk scattering configuration.

\subsection{Derivation from bulk point scattering}
\label{sec:bulkscatt}

In this subsection, we provide a simple bulk interpretation for the matching between boundary mutual informations and bulk swing surfaces in global AdS$_3$. We do so by relating the computations above to a certain forward scattering limit of the $2\rightarrow2$ processes that appear in May et al.'s~\cite{May:2019odp} connected wedge theorem.

The relevant result of \cite{May:2019odp} is that bulk $2\rightarrow 2$ scattering requires a connected entanglement wedge for certain $1\rightarrow2$ boundary ``decision regions.'' 
This result is motivated by combining holography with results about the entanglement resources needed to perform non-local quantum tasks on the boundary. 
As part of the proof, the authors show that for generic AdS$_3$ spacetimes, the mutual information of the decision regions is lower bounded by a so-called ridge surface. In this section, we will show that we can map the swing bench to this ridge surface, and the decision regions to ${\cal D}(V_i)$. Then, the fact we are in global AdS will mean the bound is saturated and reproduces
\begin{equation}
\frac{1}{2}I(V_1:V_2)=\frac{\textrm{Length}(\gamma_\textrm{swing})}{4G}\,,
\end{equation}
where for simplicity we consider a configuration for which $S(V_1)>S(X_1)$, but the opposite case can be handled analogously. 

We now explain the setup in more detail. The first step is to take the causal developments of our $V_i$ intervals in the boundary manifold:
\begin{equation}
   \mh{V}_1 = \mathcal{D}(V_1)\qquad 
   \mh{V}_2 = \mathcal{D}(V_2)\,,
    \label{inputregion}
\end{equation}
and to label the past-most tips of these diamonds as the points $c_1$ and $c_2$, respectively. The next step is to identify the two points $r_1$ and $r_2$ in the future of $c_1$ and $c_2$ obeying 
\begin{equation}
    \mh{V}_i
    =
    \hat J^+(c_i) \cap \hat J^-(r_1) \cap \hat J^-(r_2)\,, \qquad i=1,2\,.
    \label{inputt}
\end{equation}
Here, $\hat J^+(p)$ and $\hat J^-(p)$ indicate the future and past, respectively, of the point $p$, with the hats indicating a restriction to the conformal boundary, instead of including bulk points. 
One can then ask about the $2\rightarrow2$ scattering process from the $c_i$ to the $r_i$ through the bulk.

This scattering process is characterized by a ridge surface, definable in pure AdS$_3$ by 
\begin{equation}
{\cal R}_{V_1,V_2}= \partial J^+(c_1) \cap \partial J^+(c_2) \cap J^-(r_1) \cap J^-(r_2)\,,
\label{ridge}
\end{equation}
where the un-hatted $J$ here indicate bulk pasts/futures.
We are interested in cases where ${\cal R}_{V_1,V_2}$ is nonempty, indicating that bulk scattering is kinematically allowed. For general aAdS$_3$ spacetimes (and using the appropriate generalized definition of ${\cal R}_{V_1,V_2}$ from \cite{May:2021nrl}) one can then show\footnote{Technically, this result relies on mild assumptions regarding null curvatures and the maximin procedure; see \cite{May:2021nrl}.}
\be\label{ineq}
\frac{1}{2}I(V_1:V_2)\ge\frac{{\rm Length}({\cal R}_{V_1,V_2})}{4G}\,.
\ee
This appears as an intermediate step in the proof of \cite{May:2021nrl} that bulk scattering implies a connected wedge, and it arises from applying focusing arguments to various lightsheets.
In the case of global AdS, the relevant lightsheets are Killing horizons, and~\eqref{ineq} becomes an equality.

Hence, we have a bulk object that matches the mutual information that appeared in the previous subsection.
Our goal is now to argue that this ridge surface limits to the swing surface of $A'$ as $V_1$ and $V_2$ become null.

To do so, 
we define the region $A'$ with respect to $V_1$ and $V_2$ as in figure \ref{fig:regions2}.
Now,  recall that the swing surface for $A'$ is a portion of the geodesic, call it $\gamma$,
which is formed by the intersection of the two past lightcones emanating from $\partial A'_1$ and $\partial A'_2$.
Explicitly,
\begin{equation}
    \gamma_{\textrm{swing}}\subset
    \partial J^-(\partial A'_1)
    \cap 
    \partial J^-(\partial A'_2)
    \equiv \gamma \,.
\end{equation}
Meanwhile the ridge surface obeys
\begin{equation}
{\cal R}_{V_1,V_2}\subset \partial {J}^+(c_1) \cap \partial {J}^+(c_2) \equiv \gamma^{12}\,,
\end{equation}
which trivially follows from \eqref{ridge}. This equation says that the ridge surface is a portion of the geodesic $\gamma^{12}$ formed by the intersection of the future lightcones of $c_1$ and $c_2$. 
In global AdS$_3$, the future lightcone of $c_1$ is equivalent to the past lightcone of the conjugate point $c_1'$ shown in figure \ref{fig:regions2},
and similarly for $c_2$ and $c_2'$.
Moreover, in the  $\varepsilon\to 0$ limit, the points $c_2'$ and $c_1'$ approach $\partial A'_1$ and $\partial A'_2$, respectively.
Hence, we have that in the $\varepsilon\to0$ limit, $\gamma^{12}$ approaches $\gamma$.

\begin{figure}
    \centering
    \hspace{3em}
    \includegraphics[width=0.4\linewidth]{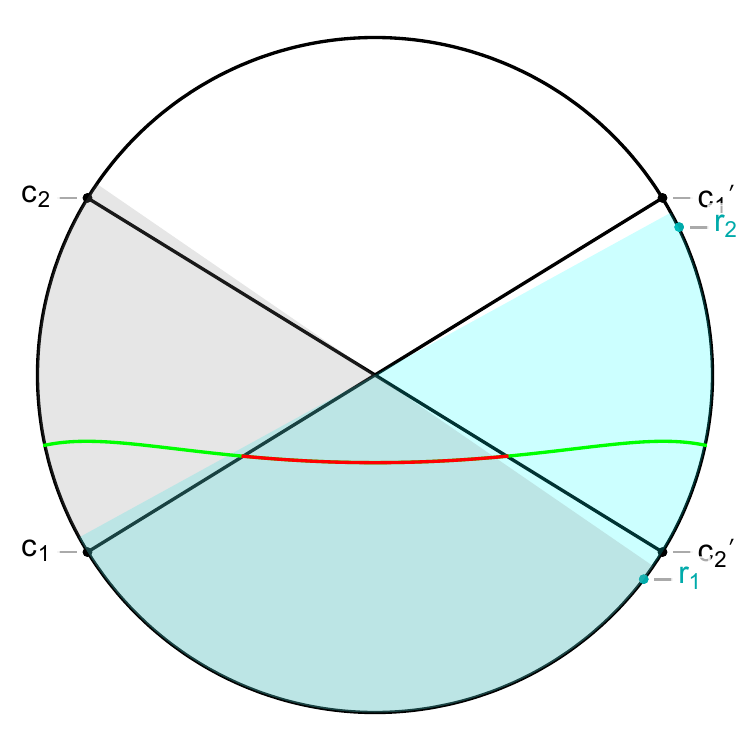}
    \qquad
    \includegraphics[width=0.45\linewidth]{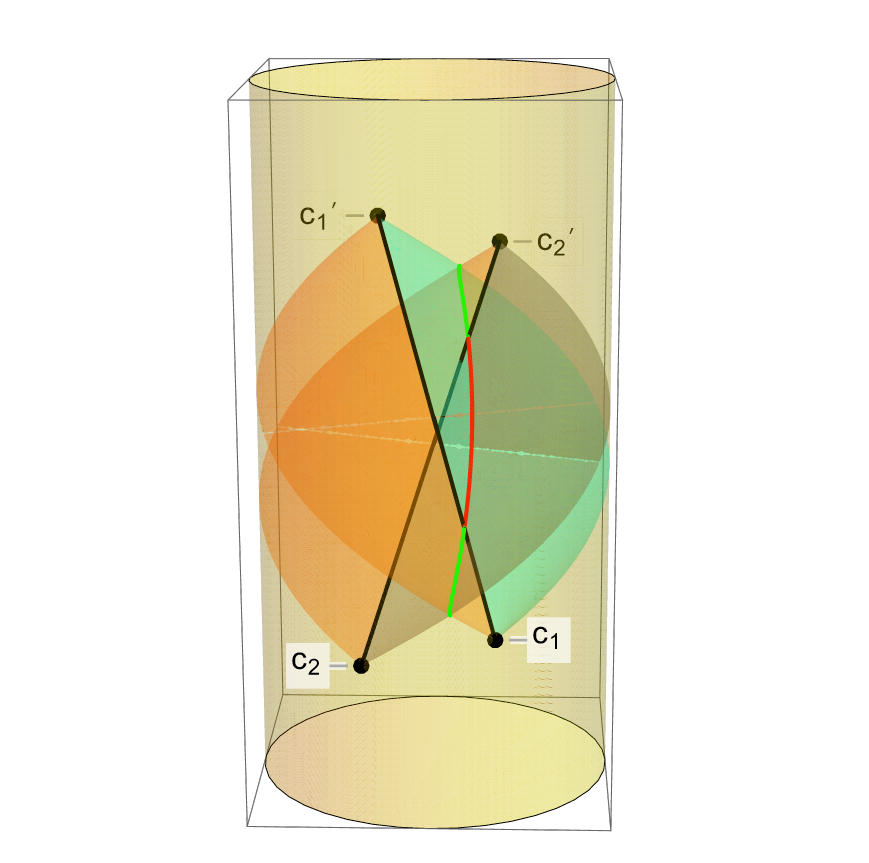}
    \caption{The ridge surface ${\cal R}_{V_1 V_2}$ is contained in the green and red geodesic $\gamma^{12}$.
    It is also contained in the blue and gray wedges defined by~\eqref{wedge1} and \eqref{wedge2}, respectively.
    In the $\varepsilon\to0$ limit, these properties ensure that the ridge surface limits to the swing surface for $A'$ (red).
    }
    \label{fig:halves}
\end{figure}

Now, in \eqref{ridge}, it is the intersections with 
$J^-(r_1)$ and $J^-(r_2)$
which delimit the ridge surface within the larger surface $\gamma^{12}$.
Meanwhile, in the definition of the swing surface, 
it is the null-rope cutoff procedure which delimits the swing surface from  the  larger surface $\gamma$.
Hence, to show that the ridge surface limits to the swing surface,
we must show that these two cutoff prescriptions agree.
To do so, note the relation 
\begin{equation}
{\cal R}_{V_1,V_2}\subset J^+(c_1) \cap J^-(r_2)
\,,
\label{wedge1}
\end{equation}
which follows from \eqref{ridge}. 
In global AdS, this relation says that the ridge surface lies within the region sandwiched between the past lightcones of $c_1'$ and $r_2$.
In the $\varepsilon\to 0$ limit, $c_1'$ approaches $r_2$, and hence these lightcones degenerate, with their intersection approaching the radial lightray running between $c_1$ and $c_1'$.
The sandwiched region is depicted in figure \ref{fig:halves}, in blue. 
Similarly, 
\begin{equation}
{\cal R}_{V_1,V_2}\subset J^+(c_2) \cap J^-(r_1)
\label{wedge2}
\end{equation}
picks out the half of the past lightcone of $c_2'$ which is shown in gray in figure \ref{fig:halves}.
Again, in the $\varepsilon\to 0$ limit, the boundary of this region approaches the radial lightray running between $c_2$ and $c_2'$.

From the appearance of these radial rays, we can conclude that ${\cal R}_{V_1V_2}$ approaches the swing surface of $A'$ in the  $\varepsilon\to 0$ limit.
This gives the desired bulk geometric interpretation for the matching between the swing formula and the mutual information in \eqref{eq:toshow}.

This construction also clarifies the apparent coordinate dependence of the swing ropes. 
In the conformal frame used here, the two chiral sectors are regulated by the same UV cutoff $\varepsilon$, and the coordinate extents \eqref{eq:VX1}, \eqref{eq:VX2} of the nearly null intervals are matched to this cutoff so as to recover $S^+$ and $S^-$ from ordinary interval entropies.
These choices cause the points $(r_1,r_2)$ to approach the points $(c_2',c_1')$ horizontally, i.e.~along slices of constant $t$ coordinate in the boundary theory;
hence, the intersections of their lightsheets are radial curves.
More generally, however, a change of conformal frame may induce distinct chiral cutoffs $\varepsilon^+$ and $\varepsilon^-$. 
With coordinate extents \eqref{eq:VX1}, \eqref{eq:VX2} of the nearly null intervals matched to these new cutoffs, 
the points $(r_1,r_2)$ no longer approach the points $(c_2',c_1')$ horizontally,
and crucially, the intersections of the limiting lightsheets are no longer radial.\footnote{Specifically,
if the displacement of two boundary points is $(\Delta \phi,\Delta \tau)=\varepsilon(\cos\theta,\sin\theta)$,
with $|\tan\theta|<1$ (spacelike separation)
then
the minimum radius of their lightcones' intersection is $r_*=\ell|\tan\theta|/\sqrt{1-\tan^2\theta}$ in the $\varepsilon\to 0$ limit.
This vanishes only for $\theta=0$ (horizontal approach).}
In this sense, the coordinate dependence of the swing ropes reflects the conformal-frame dependence of the boundary UV regulator.
We emphasize that with $\varepsilon_+\ne \varepsilon_-$, the arguments of this section and of section \ref{sec:swingMI} still go through; in particular,  both sides of the equality
\begin{equation}
    S^+(A)-S^-(A)
    =
    \frac{1}{2}I(V_1:V_2)\,
\end{equation}
simply differ from their symmetric-cutoff counterparts by the term $(c/6)\log(\varepsilon_-/\varepsilon_+)$.

To close, we emphasize that no flat limit has been taken here, only a nearly null limit of the relevant regions $V_1$ and $V_2$.
However, if we additionally take $A$ to lie within a small time band $\Delta \tau\sim 1/ \ell$, 
we find that the ridge surface probes only the Minkowski-like region with $r\ll \ell$.
Hence, the ridge surfaces of \cite{May:2019odp} are able to probe flat-space physics in global AdS$_3$, unlike usual RT surfaces.
It is interesting to ask to what extent these insights apply beyond the vacuum state, and that is the subject of the next section.

\subsection{Assumptions and limitations}
\label{sec:swingGen}

We now ask to what extent the above observations persist beyond the vacuum setting. 
First, we again emphasize that the ridge relation \eqref{ineq} need not be an equality outside of the  vacuum setting, even using the refinement of \cite{May:2021nrl}.  
That is, in typical asymptotically AdS$_3$ spacetimes,
the area of the ridge surface only lower-bounds the mutual information, rather than computing it directly. 
However, that does not necessarily rule out a duality between swing area and mutual information as in \eqref{eq:toshow}. 

Indeed, looking back at the proof in section~\ref{sec:swingMI}, we note that it technically holds not only in the vacuum but whenever the underlying state obeys the following three conditions:\footnote{A fourth would be the relation $S(V_1\cup V_2)=
\min\!\left\{
S(V_1)+S(V_2),\,
S(X_1)+S(X_2)\right\}$,
where $X_1\cup X_2$ is the complement of $V_1\cup V_2$ on the boundary Cauchy slice.
However, this relation already follows from the classical (H)RT prescription for a pure holographic state with bulk Cauchy slices of disk topology.}
the identification of $|S^+(A)-S^-(A)|$ with a swing surface area,
\begin{equation}
    \frac{\textrm{Length}(\gamma_{\textrm{swing}})}{4G}=|S^+(A)-S^-(A)|\,,
    \label{cond1}
\end{equation}
the factorization of the entanglement entropy into left- and right-moving pieces,
\begin{equation}
    S^+(A)=S(V_1)\qquad S^-(A)=S(X_1),
    \label{cond2}
\end{equation}
and purity:
\begin{equation}
    S(A) = S(\bar{A})\,.
    \label{cond3}
\end{equation}
The example at the end of section~\ref{sec:bdy-persp-2} of the spinning defect  geometries with $b_++ b_-=\ell$ nominally provides a nontrivial class of examples that satisfies the last two at least. From \eqref{facto}, we see that the entropy factorizes,
while purity is guaranteed from the topology of the geometry (and from the explicit form of \eqref{facto}).
The relation \eqref{cond1} was also discussed in section  \ref{sec:bdy-persp-2}, and is poised to falter only at the level of the minimization over swing candidates.  
As we noted below \eqref{eq:entropy-diff-spinning-defect}, the minimization step is already an issue in checks of the original swing proposal, and it would be interesting to understand if this issue can be better understood through the connection to the mutual information initiated here.
It would  also be interesting to understand to what extent other states built by chiral excitations of the vacuum obey~\eqref{eq:toshow}.\footnote{For instance, one could apply the chiral conformal unitary excitations analyzed in \cite{Caminiti:2025hjq}.
One could also consider a rotating version of the matter-shell geometry studied in~\cite{Caminiti:2025gyv},
tuned to the locus $M-M_0=|J|/\ell$.
At the level of mass and angular momentum parameters, this state also appears to represent a chiral excitation of the vacuum;
however, at the level of bulk RT surfaces, it is not entirely clear that the entropy should still factorize.
Curiously, in this example, the bulk point limit picture of section \ref{sec:basics} may require modification, since null rays emanating radially from the bulk point region will experience time delays.}

Of course, more generally, i.e. in mixed states such as black holes, 
or states without left-right factorization such as generic spinning defects,
the arguments of section \ref{sec:swingMI} will break down, and the duality between the swing surface and the mutual information will fail in its current form.
This motivates us to investigate a different approach in the following section.

\section{Swing interpretation: take II}
\label{sec:innerhor} 
In the previous section, 
we investigated whether swing surfaces in AdS$_3$ admit an interpretation in terms of boundary mutual information.
We found that this interpretation only holds subject to several strong assumptions on the underlying state,
in particular a factorization of the entanglement entropy into left- and right-moving pieces.

In this section, we pursue a different route.
Rather than attempting to identify the swing surface directly with a known entanglement measure,
we instead relate it to a black hole inner horizon by unpacking the original derivation of the swing surface. This will let us reduce the problem of finding an information-theoretic interpretation for swing surfaces to the more general problem of finding an information-theoretic interpretation for black hole inner horizons in
holography.
This connection motivates us to explore how the flat-space limit may isolate inner horizon physics already encoded in the parent CFT.
Along the way, we also gain a new perspective on the origin of the null ropes. 

To see how inner horizons connect to what we have discussed so far, let us rephrase our previous results in terms of modular flow. 
In relativistic CFTs, the entanglement entropy can be interpreted as the expectation value of an operator $K$ called the modular Hamiltonian:
\be
S(A)=-{\rm tr} \rho_A\log\rho_A=\langle K\rangle,~~~\rho_A=e^{-K}.
\ee
In the CFT vacuum state, we can split  $K=K_++K_-$ and write the uplifted swing formula as
\begin{equation}
\badat{3}
S_{swing}(A)= \tr \rho_A (K_+ - K_-)
    = \langle K_+ - K_-\rangle .
\eadat\end{equation}
In particular, the split obeys $S^+=\langle K_+\rangle$ and $S^-=\langle K_-\rangle$.
The operator $K$ generates a timelike symmetry or ``modular flow,'' while the $ K_+- K_-$ generates a spacelike symmetry. In the flat limit, the usual combination in $K$ diverges,\footnote{Explicitly, it generates a timelike conformal symmetry unsuppressed by $\ell$, and we have $\partial_\tau = \ell \partial_u\to\infty$.}  and we are seemingly left with only the spacelike flow.

These flows are closely connected to the original derivation of the swing surface in~\cite{Jiang:2017ecm}.
The point is that both linear combinations $K_+\pm K_-$ have close cousins in CFT thermal states --- and hence bulk AdS black hole geometries --- when following
the Rindler map construction of Casini-Huerta-Myers (CHM)~\cite{Casini:2011kv}. 
We describe the details of this map below, but the key feature is that in $3d$ gravity, it relates the timelike flow $K_+ + K_-$ to the  Killing symmetry generating the BTZ outer horizon and the spacelike flow  $K_+ - K_-$ to the  Killing symmetry generating the BTZ inner horizon.
Technically, the authors of~\cite{Jiang:2017ecm} focus on the flat limit of this picture, where the BTZ inner horizon becomes a flat-space cosmological horizon, but we will see here that their modified CHM construction naturally lifts to the AdS case, as also explored by \cite{Fareghbal:2020dtq,Wen:2024yny}.

In section~\ref{sec:modCHM}, we follow a Rindler map route to clarify the connection to inner horizons, as well as the role of the radial cutoffs in the swing prescription.
These results motivate the search for measures of inner horizon thermodynamics in the AdS uplift.
This is then complemented by the results in section~\ref{sec:subCardy}, where we demonstrate how the inner horizon appears to contribute as a subleading saddle in the Cardy formula that survives as the dominant saddle in the flat limit.

\subsection{Uplifting the modified CHM construction}
\label{sec:modCHM}

In this section, we approach the problem of interpreting swing surfaces through a connection to inner horizon physics.
We do this via a modification of the Rindler map construction in Casini-Huerta-Myers~\cite{Casini:2011kv}.
We reverse the logic of CHM, in that we use a proposed subregion entanglement measure to motivate looking for a thermodynamic interpretation of the inner horizon,
instead of using known horizon physics to prove an entanglement entropy formula.

Our starting point is the standard thermodynamic interpretation of the outermost black hole horizon in AdS/CFT. 
In the semiclassical regime, the Bekenstein-Hawking entropy of an asymptotically AdS black hole agrees, at leading order in $G$, with the thermal entropy of the dual boundary state \cite{Bekenstein:1973ur,Hawking:1975vcx,Witten:1998zw}. 
In AdS$_3$, this relation takes the form \cite{Strominger:1997eq}
\begin{equation}
S(\rho_\beta)
=
\frac{\operatorname{Length}(\gamma_{\mathrm{out}})}{4G}\,,
\label{bh}
\end{equation}
where $\gamma_{\mathrm{out}}$ is a spatial cross-section of the outer horizon,
$\rho_\beta$ is the corresponding equilibrium state of the boundary CFT,
and $S$ is the von Neumann entropy. 
For a rotating, charged, or otherwise generalized black hole with multiple horizons,
we will use the symbol $\beta$ as shorthand for the full collection of thermodynamic potentials, such as $(\beta,\Omega,\mu,\ldots)$.

Suppose now that the bulk solution also contains an inner horizon $\gamma_{\mathrm{in}}$. 
Our central assumption is that its length likewise computes some quantity intrinsic to the dual microscopic state. Denoting this quantity by $S_*$, we are assuming that
\begin{equation}
S_*(\rho_\beta)
=
\frac{\operatorname{Length}(\gamma_{\mathrm{in}})}{4G}\,.
\label{kl}
\end{equation}
In the following section, we will argue that a quantity similar to $S_*$ can be realized as a subleading contribution to a microcanonical thermodynamic entropy in AdS/CFT (see also recent work~\cite{Kruthoff:2024gxc,Caminiti:2026efx}), and we explore more options in the discussion section.
For now, however, we will remain agnostic to the precise definition of $S_*$.
The only structural property we require  is invariance under unitary conjugation,
\begin{equation}
S_*(U\rho U^\dagger)=S_*(\rho)\,,
\end{equation}
which many information-based quantities naturally satisfy. 
Our claim is that, if such a boundary interpretation of the inner horizon exists, then a corresponding interpretation of the swing surface follows.

The mechanism underlying this claim is closely analogous to the construction of Casini, Huerta, and Myers (CHM) in \cite{Casini:2011kv}. 
We therefore begin by reviewing that construction, first schematically, and then in much more detail.

\paragraph{Review of the Rindler map}
Consider a spatial interval $A$ in the vacuum state of a two-dimensional CFT.
Its reduced density matrix $\rho_A$ is uniquely characterized by its modular flow,
namely the Heisenberg evolution generated by its modular Hamiltonian  $K=-\log\rho_A$.
This flow is constrained by conformal symmetry to have a very simple, almost thermal, form.
In particular, if one conformally maps the causal development $D(A)$ of the interval to a complete copy of Minkowski spacetime,\footnote{Note that in higher dimensions, the role of $\mathbb{R}^{1,1}$ is played by a hyperbolic space, $\mathbb{R}\times H^{d-1}$.}
\begin{equation}
D(A)
\rightarrow
\mathbb{R}^{1,1}\,,
\label{confg}
\end{equation}
then the vacuum modular flow on $D(A)$ is mapped to ordinary time translation on $\mathbb{R}^{1,1}$.
It follows that the reduced density matrix $\rho_A$ is conformally equivalent to a thermal state with respect to the Hamiltonian generating these time translations. 
At the level of appropriately regulated density matrices, one may write
\begin{equation}
\rho_A
=
U\rho_\beta U^\dagger,
\label{conf}
\end{equation}
where $U$ implements the conformal transformation, and the inverse temperature $\beta$ is fixed by the periodicity of the Euclidean modular flow.
By the invariance of the von Neumann entropy under unitary conjugation, 
it also follows that we have
\begin{equation}
S(\rho_A)=S(\rho_\beta)\,.
\end{equation}
Hence, the conformal map \eqref{confg} turns the problem of computing a subregion entanglement entropy $S(\rho_A)$ into the problem of computing a thermodynamic entropy $S(\rho_\beta)$.

Now, in a holographic CFT$_2$, $\rho_\beta$ has a direct gravitational interpretation.
To find the bulk geometry associated to $\rho_\beta$,
one notes that the boundary conformal transformation \eqref{confg} extends into the bulk as a coordinate transformation. 
Under this transformation, the bulk region associated with $D(A)$ becomes the AdS-Rindler spacetime.
If we interpret this spacetime as a topological black hole, then \eqref{bh} gives 
\begin{equation}
    S(\rho_\beta)
=
\frac{\operatorname{Length}(\gamma_{\mathrm{out}})}{4G}\,,
\end{equation}
where here $\gamma_{\mathrm{out}}$ denotes the AdS-Rindler horizon.
Crucially,
when expressed in the original vacuum coordinates, 
the AdS-Rindler horizon is precisely the RT surface $\gamma_A$ anchored at the endpoints of $A$. 
That is, with compatible regulators understood, one has
\begin{equation}
\operatorname{Length}(\gamma_A)
=
\operatorname{Length}(\gamma_{\mathrm{out}}).
\end{equation}
The relevant result of CHM is that combining the previous three equations gives a complete derivation of the RT formula:
\begin{equation}
S(\rho_A)
=
S(\rho_\beta)
=
\frac{\operatorname{Length}(\gamma_{\mathrm{out}})}{4G}
=
\frac{\operatorname{Length}(\gamma_A)}{4G}\,.
\end{equation}
The first equality follows from the conformal equivalence of the two states, the second from black hole thermodynamics, and the third from the bulk extension of the conformal map.

\paragraph{Modified Rindler map}
For our purposes,  the CHM construction provides more than just a derivation of holographic entanglement entropy. 
It reveals that, in some sense, vacuum RT surfaces inherit their boundary interpretation from the Bekenstein-Hawking formula for black hole horizons. Our goal is to achieve a similar result for uplifted AdS swing surfaces: namely that they inherit the information content of inner horizons.
Schematically, the formula we are after is 
\begin{equation}
S_*(\rho_A)
=
S_*(\rho_\beta)
=
\frac{\operatorname{Length}(\gamma_{\mathrm{in}})}{4G}
=
\frac{\operatorname{Length}(\gamma_{\textrm{swing}})}{4G}\,.
\label{desrrr}
\end{equation}
To obtain this result, we will perform a small modification of the conformal map in CHM,
and then repeat the CHM procedure in detail.
As in CHM, we begin with the standard reduced density matrix for the vacuum state.
For concreteness, we assume that $A$ is an interval in Minkowski spacetime $\mathbb{R}^{1,1}$ with metric
\begin{equation}
    ds^2=dx_+ dx_- = -dt^2+dx^2\,.
    \label{mink}
\end{equation}
We take $D(A)$ to have extent $(\Delta x_+$, $\Delta x_-)$ in the null directions.
Then, the modular Hamiltonian $K=-\log \rho_A$ generates the following timelike conformal symmetry in the diamond \cite{Casini:2011kv,Hislop:1981uh}:
\begin{equation}
\xi_A
=
2\pi
\left[
\frac{\Delta x_+^2/4-x_+^2}
{\Delta x_+}\,\partial_{+}
-
\frac{\Delta x_-^2/4-x_-^2}
{\Delta x_-}\,\partial_{-}
\right].
\label{eq:modflow}
\end{equation}
Following \cite{Jiang:2017ecm}, we introduce two parameters $R_+$ and $R_-$ with $R_+>R_->0$,
together with the following conformal map from $x_\pm$ coordinates to new $X_\pm$ coordinates: 
\begin{equation}
   X_\pm
   =
   \frac{\beta_\pm}{\pi}\textrm{arctanh}\frac{x_\pm}{\Delta x_\pm/2}\,,
   \qquad
   \beta_\pm = \frac{2\pi}{R_+\pm R_-}\,.
   \label{newconf}
\end{equation}
This map still has the effect of mapping $D(A)$ to $\mathbb{R}^{1,1}$ as in \eqref{confg},
however, it is different from 
the conformal map of CHM;
in CHM, the authors effectively set $R_-=0$ (see (2.25) of \cite{Casini:2011kv}).

We now investigate the state $\rho_\beta$ which results from performing this new conformal map.
On the $X_\pm$ plane with metric
\begin{equation}
    ds^2 = dX^+ dX^- = -dT^2 + dX^2\,,
\end{equation}
the modular flow in \eqref{eq:modflow} becomes 
\begin{equation}
    \xi_\beta =\beta\left(\partial_T-\Omega \partial_X\right)\,,
    \qquad
    \beta\equiv \frac{2\pi R_+}{R_+^2-R_-^2}\,,
    \qquad
    \Omega\equiv  \frac{R_-}{R_+}\,.
    \label{xi-beta}
\end{equation}
To interpret this result, let us write the generator of time translations as $H$ and the generator of spatial translations as $P$.
Then, it is clear that  the above equation equates $-\log \rho_\beta$ with $\beta(H-\Omega P)+\log Z$.
That is, our new state 
$\rho_\beta$ is the  density matrix for a 
thermal state on $\mathbb{R}^{1,1}$,
with a nontrivial chemical potential $\Omega$ for linear momentum.
Again, here the main difference to CHM is that CHM would set $R_-=0=\Omega$ and hence obtain an ordinary thermal state.

Still, we can follow the logic of CHM, and the next step is to identify the bulk extension of the conformal map \eqref{newconf}.
In this case, it reads\footnote{Technically, this map only does the job for $R>R_+$. 
With $R_-<R<R_+$, 
one must replace $\sqrt{R^2-R_+^2} \textrm{ch} v$ 
by 
$\sqrt{R_+^2-R^2} \textrm{sh} v$ 
and similarly, replace
$\sqrt{R^2-R_+^2} \textrm{sh} v$ 
by 
$\sqrt{R_+^2-R^2} \textrm{ch} v$,
where we define $v=R_+T+R_-X$.
This result can be derived by going from Poincar\'e AdS  to rotating Rindler AdS through embedding space.
\label{behind}
}
\begin{equation}
\begin{aligned}
\frac{x_+}{\Delta x_+/2}
&=
\frac{
\sqrt{R^2-R_-^2}\textrm{sh} (R_-T+R_+X)
+
\sqrt{R^2-R_+^2}\textrm{sh} (R_+T+R_-X)
}{\mathcal D},
\\
\frac{x_-}{\Delta x_-/2}
&=
\frac{
\sqrt{R^2-R_-^2}\textrm{sh} (R_-T+R_+X)
-
\sqrt{R^2-R_+^2}\textrm{sh} (R_+T+R_-X)
}{\mathcal D},
\\
\frac{z}{\sqrt{\Delta x_+ \Delta x_-/4}}
&=
\frac{
\sqrt{R_+^2-R_-^2}
}{\mathcal D},
\end{aligned}
\label{eq:bulk-rotating-rindler-map}
\end{equation}
where we have shortened $\cosh$ and $\sinh$ to $\textrm{ch}$ and $\textrm{sh}$ and defined
\begin{equation}
\mathcal D
\equiv
\sqrt{R^2-R_-^2}\textrm{ch} (R_-T+R_+X)
+
\sqrt{R^2-R_+^2}\textrm{ch} (R_+T+R_-X).
\end{equation}
This transformation relates the ordinary Poincar\'e AdS$_3$ metric
\begin{equation}
   ds^2 = \frac{dx_+dx_-+dz^2}{z^2}\,,\qquad \ell\equiv 1\,,
\end{equation}
which is  dual to the vacuum state on \eqref{mink},
to the so-called rotating Rindler AdS$_3$ spacetime.
This spacetime is described by the following metric \cite{Jiang:2017ecm,Parikh:2011aa}
\begin{equation}
ds^2 = 
-
\frac{(R^2-R_+^2)(R^2-R_-^2)}{R^2} dT^2
+
\frac{ R^2 dR^2}{(R^2-R_+^2)(R^2-R_-^2)} 
+ R^2 \left(dX + \frac{R_+ R_-}{R^2} dT\right)^2\,,
\label{eq:btz-metric-rpm}
\end{equation}
with $R>0$ and $T,X \in \mathbb{R}$.
It has both an outer horizon at $R=R_+$
and an inner horizon at $R=R_-$.
Note that while it is conventional to call this spacetime ``rotating,'' this is a bit of a misnomer, since the $X$ direction is infinite rather than periodic.
Also note that with $R_-=0$ we recover the ordinary Rindler-AdS$_3$ spacetime considered in CHM.
We now see why it has been useful to  modify the conformal map of CHM, since it allows us to introduce a nontrivial inner horizon in the bulk spacetime.

Before discussing inner horizons, however,
let us reassure ourselves that we can still derive the RT formula by interpreting this new Rindler spacetime as our topological black hole.
To derive the RT formula, the first step is to appropriately regularize the density matrix $\rho_A$.
Following \cite{Jiang:2017ecm}, we will place UV cutoffs on $\rho_A$ by introducing null lattice spacings $\epsilon_{\pm}$ in spacetime.
These lattice spacings regulate the entanglement between $A$ and $\bar{A}$ by eliminating modes of arbitrarily high frequency which
otherwise straddle the boundary of $A$ and cause infinite entanglement with $\bar{A}$.
In other words, their primary  role is to exclude the QFT degrees of freedom in the tiny diamond of spacetime $D_{++}$ defined as follows:
\begin{equation}
    D_{++} \equiv \left\{ x_+ \in \left(\frac{\Delta x_+}{2}-\epsilon_+, \frac{\Delta x_+}{2}+\epsilon_+\right) \textrm{    and    } 
    x_- \in \left(\frac{\Delta x_-}{2}-\epsilon_-, \frac{\Delta x_-}{2}+\epsilon_-\right) \right\}
    \,,
    \label{excludeme}
\end{equation}
near the endpoint of $A$ at $(x_+,x_-)=(\Delta x_+/2,\Delta x_-/2)$,
and analogously for a tiny diamond $D_{--}$ near the other endpoint at 
$(-\Delta x_+/2,-\Delta x_-/2)$.

\begin{figure}
    \centering
 
    \includegraphics[width=0.8\linewidth]{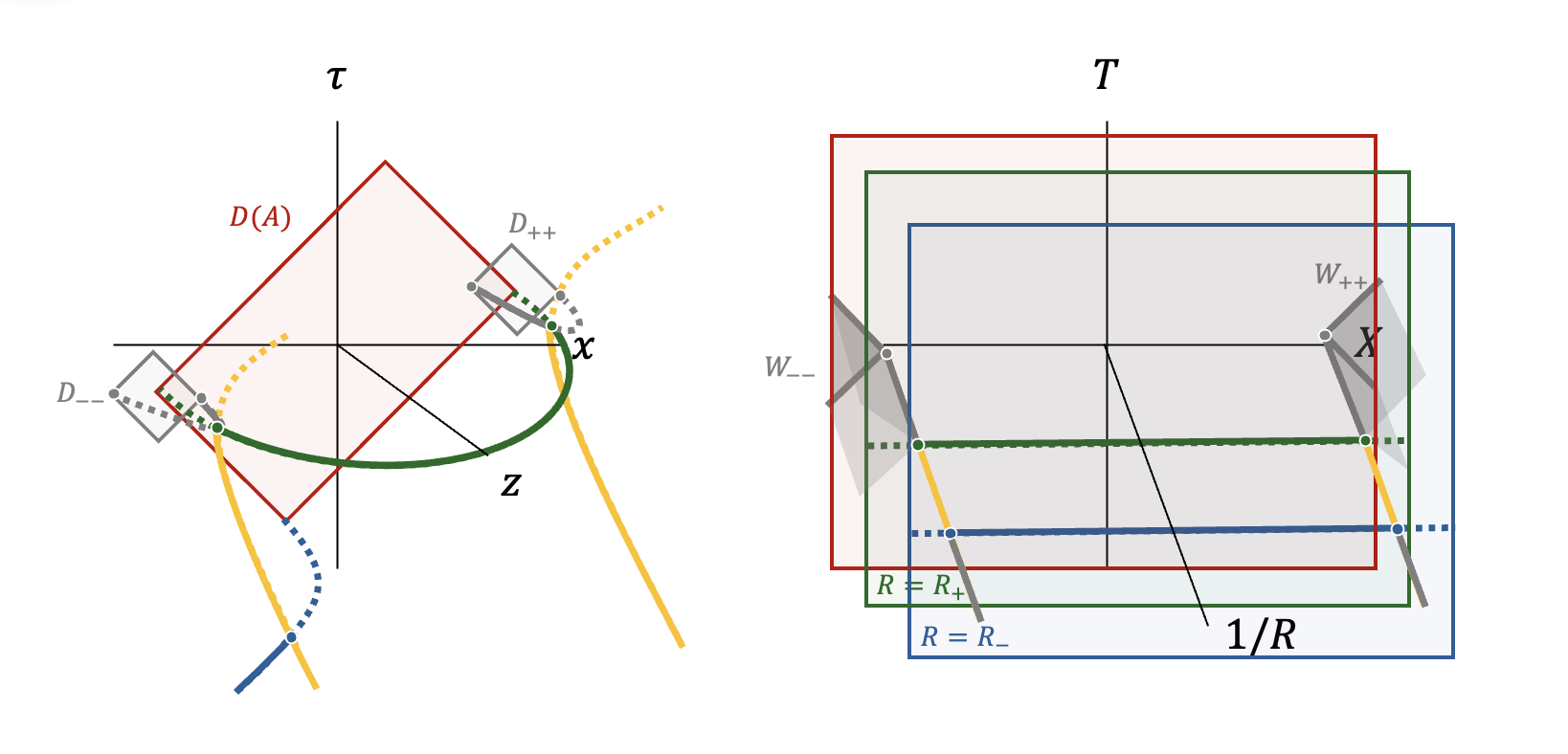}
    \caption{
    The RT surface in Poincar\'e coordinates (green) is mapped to the outer horizon, $R=R_+$, in AdS-Rindler coordinates (green).
    The swing surface in Poincar\'e coordinates (blue) is mapped to the inner horizon, $R=R_-$, in AdS-Rindler coordinates (blue). 
    }
    \label{fig:not-null}
\end{figure}

Removing these boundary cutoff regions corresponds, in the bulk, to removing their associated entanglement wedges.
So, a convenient bulk realization of the boundary cutoffs is supplied by their two RT surfaces.
The RT surface for $D_{++}$ reads
\begin{equation}
\begin{cases}
\frac{x_+ - \Delta x_+/2}{\epsilon_+}
=
\frac{x_- - \Delta x_-/2}{\epsilon_-}
,\\
\left(x_+ - \frac{\Delta x_+}{2}\right)
\left(x_- - \frac{\Delta x_-}{2}\right)
+z^2
=\epsilon_+ \epsilon_-\,.
\end{cases}
\label{eq:cutoff-curve-quadratic}
\end{equation}
The equation for $D_{--}$ follows by reversing the endpoint signs.
These cutoff RT surfaces are shown as the tiny gray curves in figure \ref{fig:not-null}, left.
Equivalently, one may impose the usual near-boundary cutoff $z=\sqrt{\epsilon_+\epsilon_-}$; this avoids circularly using the RT prescription itself in deriving the RT formula.

The next step is to 
track how these regulators behave under the boundary conformal map \eqref{newconf} which relates $\rho_A$ to $\rho_\beta$.
From \eqref{newconf}, one finds that the UV cutoffs $\epsilon_\pm$ in the boundary $x_\pm$ plane translate to IR cutoffs  $\Delta X_{\pm}$ on the $ X_\pm$ plane, as follows:
\begin{equation}
   \Delta X_{\pm}
    = 
    \frac{\beta_{\pm}}{\pi}
    \log \frac{\Delta x_{\pm}}{\epsilon_{\pm}}\,.
    \label{cutoffs}
\end{equation}
That is, in the $X_\pm$ plane, 
the excluded $D_{++}$ region corresponds to the following infinite and causally complete region of spacetime:
\begin{equation}
    W_{++} = \left\{ X_+>\Delta X_+/2 \textrm{   and   } X_->\Delta X_-/2\right\}, 
    \label{exclude}
\end{equation}
and similarly for $W_{--}$. Their bulk regulator curves are
\begin{equation}
    (X_+, X_-) = \left(\pm \frac{\Delta X_+}{2},\pm\frac{\Delta X_-}{2}\right)
    \,,
    \qquad R>R_+\,,
    \label{cutcurve}
\end{equation}
shown as the straight, solid gray lines in figure \ref{fig:not-null}.
Under the bulk map \eqref{eq:bulk-rotating-rindler-map}, they  reproduce \eqref{eq:cutoff-curve-quadratic} to leading order in the cutoffs.

With these regulators understood,
we may now write the unitary equivalence relation $U^\dag\rho_A U = \rho_\beta$
and the resulting entropy equivalence $S(\rho_A)=S(\rho_\beta)$.
To then derive the RT formula, we remind ourselves that
$S(\rho_\beta)$  is given by the regulated length of the outer horizon  in \eqref{eq:btz-metric-rpm}. 
But in Poincar\'e coordinates, the outer horizon $R=R_+$  is the surface
\begin{equation}
    x_+ x_- + z^2 = \frac{\Delta x_+\Delta x_-}{4}\,,
    \qquad
    \frac{x_+}{\Delta x_+} = \frac{x_-}{\Delta x_-}\,,
    \label{rtsurfer}
\end{equation}
which is precisely the RT surface for $A$.
This proves the RT formula, up to checking the cutoffs.

The cutoff surfaces intersect the horizon, or equivalently the RT surface, at 
\begin{equation}
    z_*=\frac{\sqrt{\epsilon_+\epsilon_-}}{1+\frac{\epsilon_+\epsilon_-}{\Delta x_+ \Delta x_-}}\,.
\end{equation}
Computing the length of the RT surface with these cutoffs in place
gives the expected formula for the entanglement entropy,
\begin{equation}
    S(\rho_A)
    =
    \frac{\ell}{4G}\log\frac{\Delta x_+\Delta x_-}{\epsilon_+ \epsilon_-}\,,
\end{equation}
which concludes the derivation.
We emphasize that the final answer for the RT surface does not depend on the choices of $R_+$ and $R_-$ parameters appearing in the conformal transformation \eqref{newconf},
which was our main departure from the original work of CHM.
The upshot is that without loss of generality, we can always view the RT surface as arising from an outer horizon of a geometry with two horizons ($R_+>R_->0$).

We are finally prepared to discuss inner horizons and  their relation with swing surfaces.
Recall that our central assumption in this section is the existence of a unitarily invariant boundary dual \eqref{kl} for inner horizon area, called $S_*$. 
In this case, the work we have done so far already establishes the first two equals signs in \eqref{desrrr}, namely
\begin{equation}
    S_*(\rho_A)=S_*(\rho_\beta) = \frac{\textrm{Length}(\gamma_{\textrm{in}})}{4G}\,,
\end{equation}
where $\gamma_{\textrm{in}}$ is the inner horizon, regulated by the cutoff surfaces in \eqref{cutcurve}.
It remains only to identify $\gamma_{\textrm{in}}$ with the swing surface of $A$. 

At first glance, this seems straightforward.
Using the coordinate transformation \eqref{eq:bulk-rotating-rindler-map},
the inner horizon $R=R_-$ maps to the following surface in Poincar\'e coordinates:
\begin{equation}
    x_+ x_- + z^2 = -\frac{\Delta x_+\Delta x_-}{4}\,,
    \qquad
    x_+  \Delta x_-  = -x_-  \Delta x_+\,.
\end{equation}
This curve is the intersection of the past lightcones emanating from $\partial A_1$ and $\partial A_2$.
Hence from the discussion of section \ref{sec:bulk-persp-2}, we know that it contains the swing surface.

However, from here, the regularization procedure is very subtle.
Note that so far we have only described the regulator curves  in \eqref{cutcurve}
for $R>R_+$, 
where they correspond to spacelike surfaces extending from the asymptotic boundary to the RT surface for $A$.
On the other hand, the region $R<R_+$ is very different.
Consulting the metric \eqref{eq:btz-metric-rpm}, we see that when passing behind the outer horizon, the regulator curves \eqref{cutcurve} change signature from spacelike to timelike --- it is, overall, a complex geodesic!\footnote{The real continuation of a constant-$(X_+,X_-)$ regulator through the
outer bifurcation surface enters another exterior, rather than the
black hole interior. Indeed, choosing the affine parameter $s$ to
vanish at the bifurcation surface, locally one would have
$R(s)=R_++\alpha^2s^2+\cdots$, with $\alpha^2>0$.
The interior segment is instead obtained by analytically continuing
$s=i\sigma$, giving $R=R_+-\alpha^2\sigma^2+\cdots$.}
This change of causal character is indicated by the gray (spacelike) vs.~yellow (timelike) coloring in figure \ref{fig:not-null}, right.
Using the appropriate change of coordinates discussed in footnote \ref{behind},
the continuation of the $D_{--}$ cutoff curve past the RT surface is specified by the following timelike hyperbola:
\begin{equation}
\begin{cases}
\frac{x_+ + \Delta x_+/2}{\epsilon_+}
=
-
\frac{x_- + \Delta x_-/2}{\epsilon_- }
,\\
\left(x_+ + \frac{\Delta x_+}{2}\right)
\left(x_- + \frac{\Delta x_-}{2}\right)
+z^2
=\epsilon_+ \epsilon_-\,,
\end{cases}
\label{eq:cutoff-curve-quadratic-2}
\end{equation}
and similarly for the $D_{++}$ cutoff curve.
We comment that in fact the curve is radial when $\epsilon_+$ and  $\epsilon_-$ are equal, and becomes null 
in the limit $\epsilon_\pm\to 0$ at fixed ratio $\epsilon_+/\epsilon_-$.
Hence, the radial null ray prescription can again be understood as arising from a choice of UV cutoffs in the boundary theory.

The cutoff surfaces reach the inner horizon, or equivalently the  swing surface, at
\begin{equation}
    z_*
    = \frac{\Delta x_+ \Delta x_-}{|\Delta x_+\sqrt{\frac{\epsilon_-}{\epsilon_+}}
    -
    \Delta x_-\sqrt{\frac{\epsilon_+}{\epsilon_-}}|}\,.
\end{equation}
Only one of the regulator surfaces actually hits the swing surface in the Poincar\'e patch; the second  regulator surface hits the swing surface after exiting the Poincar\'e patch.
After taking this into account, the total length of the swing surface is
\begin{equation}
    S_*(\rho_A) =
    \frac{\ell}{4G} \log\frac{\Delta x_+}{\Delta x_-}
    +
    \frac{\ell}{4G}
    \log\frac{\epsilon_-}{\epsilon_+}\,,
\end{equation}
matching \cite{Jiang:2017ecm}.

\vspace{1em}\noindent
We conclude that the existence of an information-theoretic dual for inner horizon areas would imply the existence of an information-theoretic dual for swing surface areas in the AdS uplift. We note that the manipulations performed here are essentially the AdS uplift of the flat-space derivations in~\cite{Jiang:2017ecm}, where we replace assumptions about the spacelike ``modular'' flow with assumptions about inner horizons in AdS/CFT.
In the process, we saw how the radial ropes appearing in the swing construction emerge naturally from a choice of UV cutoffs in the boundary CFT. 
In the final subsection, we come full circle  by studying inner horizons in AdS/CFT.

\subsection{Subleading Cardy saddles and the inner horizon}
\label{sec:subCardy}

What CFT observables can play the role of the bulk quantity $S_*$ proposed in \eqref{kl}?
To understand this, we can actually turn to the AFS/BMSFT literature.
Besides the swing proposal,
another main success of this program  
has been the matching of the thermal entropy of flat-space cosmologies (FSC) to a boundary Cardy computation~\cite{Barnich:2012xq,Bagchi:2012xr}. 
As mentioned above, this result underlies the swing proposal in~\cite{Jiang:2017ecm} via CHM techniques. 
In the previous section, we essentially uplifted the arguments of
\cite{Jiang:2017ecm} 
to AdS, where the FSC horizon crucially becomes the BTZ inner horizon.

Motivated by this connection,
we now argue that the BTZ inner horizon contributes a subleading saddle in the Cardy \cite{Cardy:1986ie} density of states in the dual CFT$_2$, as initially suggested in \cite{Bagchi:2012xr}.
Of course, this is a microcanonical quantity, so it cannot play the role of the canonical-ensemble quantity $S_*$ (see further comments in the discussion section).
Nevertheless, this result provides nontrivial evidence that inner horizon physics can be encoded as an information-theoretic observable in the dual CFT.

We start with a theory on the cylinder with coordinates $y^\pm = \phi \pm t$, and the usual Wick rotation $w=iy^+$ and $\bar{w}=-iy^-$. On this cylinder,
consider the CFT$_2$
 partition function 
\begin{equation}\label{eq:zdfirst}
    Z_0(\tau,\bar{\tau}) 
    =
    \tr e^{2\pi i \left[(L_0-\frac{c}{24}) \tau-(\bar{L}_0-\frac{\bar{c}}{24})\bar{\tau}\right]}\,.
\end{equation}
When we recognize $L_0=-\partial_w$ and $\bar{L}_0=-\partial_{\bar{w}}$, we see that the trace implements the joint complex identification
\begin{equation}
  \begin{pmatrix}
        w\\
        \bar{w}
    \end{pmatrix}
    \sim
    \begin{pmatrix}
        w+2\pi i\tau\\
        \bar{w}-2\pi i \bar{\tau}
    \end{pmatrix}
\end{equation}
while the original relation $\phi\sim\phi+2\pi$ gives
$(w,\bar{w})\sim (w+2\pi i,\bar{w}-2\pi i)$.
When $\tau$ and $\bar{\tau}$ are purely imaginary,
with $\tau=i\beta_+/2\pi$ and $\bar{\tau}=-i\beta_-/2\pi$,
\eqref{eq:zd} describes the usual canonical ensemble  partition function with temperature $\beta$ and angular potential $\Omega$, with $\beta_\pm \equiv \beta(1\mp \Omega)$ and the complex thermal identification $(t,\phi)\sim(t+i\beta,\phi-i\beta\Omega)$.

From \eqref{eq:zdfirst}, the density of states $D(h,\bar{h})$ is related to the partition function via 
\begin{equation}\label{eq:zd}
    Z_0(\tau,\bar{\tau})=
    \sum D(h, \bar{h}) e^{2\pi i \left[\tau E-\bar{\tau}\bar{E}\right]}\,,
    \qquad 
    E \equiv h-\frac{c}{24}\,, \qquad \bar{E} \equiv \bar{h}-\frac{\bar{c}}{24}\,.
\end{equation}
We would like to constrain $D(h, \bar{h})$ using modular invariance. 
Following \cite{Carlip:1998qw}, we strip off simple prefactors by introducing $Z(\tau,\bar{\tau})$ as follows:
\begin{equation}\label{z0}
    Z_0(\tau,\bar{\tau}) = Z(\tau,\bar{\tau})e^{-i\frac{\pi}{12}  (\tau c-\bar{\tau}\bar{c})}
    \qquad
    Z(\tau,\bar{\tau}) 
    = \tr\left[e^{2\pi i(\tau L_0-\bar{\tau} \bar{L}_0)}\right]\,.
\end{equation}
The modular invariance condition $Z_0(\tau,\bar{\tau})=Z_0(-1/\tau, -1/\bar{\tau})$ 
then becomes
\begin{equation}\label{eq:modinv}
    Z(\tau,\bar{\tau}) 
    = e^{i \frac{\pi}{12}(c\tau-\bar{c}\bar{\tau})}
    e^{i\frac{\pi}{12}\left(\frac{c}{\tau}-\frac{\bar{c}}{\bar{\tau}}\right)}Z(-1/\tau,-1/\bar{\tau})\,.
\end{equation}
Now we can invert the relationship between $Z$ and $D$ to get
\begin{equation}\label{eq:phasefactor}
    D(h, \bar{h}) = 
    \int   d\tau d\bar{\tau} e^{-2\pi ih\tau}e^{2\pi i\bar{h}\bar{\tau}}
    Z(\tau,\bar{\tau})\,
\end{equation}
for an appropriate choice of inverse-Laplace contour. Combining this with the modular invariance statement~\eqref{eq:modinv} gives
\begin{equation}
    D(h,\bar{h}) =  \int d\tau d \bar{\tau} 
    e^{2\pi i \left(-E \tau + \frac{c}{24\tau}\right)}
    e^{2\pi i \left(\bar{E} \bar{\tau} - \frac{\bar{c}}{24\bar{\tau}}\right)}
    Z\left(-1/\tau,-1/\bar{\tau}\right)\,.
    \label{negtemp}
\end{equation}
At large weights $h,\bar{h}\gg c,\bar{c}$, it is straightforward to argue that a saddle-point approximation will be valid with $Z\left(-1/\tau,-1/\bar{\tau}\right)$ slowly varying, 
so that  $D(h,\bar{h})$ is a sum over saddle points, determined by evaluating extrema of the phase. 

The usual Cardy formula then comes from evaluating the leading saddle. 
One finds that the asymptotic density of states grows like
\begin{equation}
     D(h, \bar{h})\equiv e^{S(h, \bar{h})}
    \sim \exp\left[
    2\pi\left(\sqrt{\frac{c E}{6}} + \sqrt{\frac{\bar{c} \bar{E}}{6} }\right)\right]=\exp\left[ \frac{{\cal A}_{\textrm{out}}}{4G}\right]\,,
    \label{eq:dos}
\end{equation}
where ${\cal A}_{\textrm{out}}\equiv \textrm{Length}(\gamma_{\textrm{out}})$ is the length of the BTZ outer horizon in the bulk, using the dictionary
\be
c=\bar c=\frac{3\ell}{2G},\qquad
E=\frac{\ell M+J}{2},
\qquad
\bar E=\frac{\ell M-J}{2},
\ee
where we are assuming Einstein gravity and that the Cardy formula extrapolates to the case of holographic (i.e. large) central charges, as is justified under certain assumptions by the work of \cite{Hartman:2014oaa}.

However, this is not the only saddle. In total, there are  \textit{four} saddles which have the potential to contribute:
\begin{equation}
\begin{cases}
(\tau, \bar{\tau}) =  \left(\pm i \sqrt{\frac{c}{24E}},
\pm i \sqrt{\frac{\bar{c}}{24\bar{E}}}\right)\\
(\tau, \bar{\tau}) =  \left(\pm i \sqrt{\frac{c}{24E}},
\mp i \sqrt{\frac{\bar{c}}{24\bar{E}}}\right)\,.
\end{cases}
\label{eq:4saddles}
\end{equation}
Besides the leading saddle giving~\eqref{eq:dos}, 
the next-to-largest saddle has
\begin{equation}\label{submicro}
 D_{sub}(h,\bar{h})\sim \exp\left[ 2\pi\left| \sqrt{\frac{cE}{6}}-\sqrt{\frac{\bar{c}\bar{E}}{6}}\right|\right]=\exp\left[\frac{{\cal A}_{\textrm{in}}}{4G}\right],
\end{equation}
corresponding to the microcanonical entropy of the inner horizon (see also~\cite{Castro:2012av,Detournay:2012ug,Bagchi:2012xr}). 
Also including the other saddles gives the following possible expansion for the BTZ density of states:
\begin{equation}\label{nonpert0}
    e^{S(h,\bar{h})} 
    \sim \# \exp\left[\frac{\mathcal{A}_{\textrm{out}}}{4G}\right]
    +
    \# \exp\left[\frac{\mathcal{A}_{\textrm{in}}}{4G}\right]
    +
    \# 
    \exp\left[-\frac{\mathcal{A}_{\textrm{in}}}{4G}\right]
    +
    \# 
    \exp\left[-\frac{\mathcal{A}_{\textrm{out}}}{4G}
    \right]\,,
\end{equation}
where the $\#$ incorporate loop effects.

It is now natural to worry that accessing these subleading saddles requires evaluating the integrand in \eqref{negtemp} at negative temperatures $\beta_\pm<0$, where the exact partition function becomes ill-defined due to a diverging thermal trace.
However, we emphasize that we are analytically continuing to negative temperatures  not for the exact partition function but instead for the partition function in the Cardy approximation.
In any case, we can motivate \eqref{nonpert0} in a concrete example. 
A suggestive version of this formula comes from considering the standard  vacuum entry of the Virasoro modular $S$-kernel (see, e.g.~\cite{Collier:2019weq}):
\begin{equation}
    D_{0}(P,\bar{P})
 \equiv(4\sqrt{2})^2\,
 \sinh\!\left(\frac{2\pi P}{b}\right)
 \sinh\!\left(\frac{2\pi\Pbar}{b}\right)
 \sinh(2\pi bP)\sinh(2\pi b\Pbar),
\end{equation}
where
\begin{equation}
 c=1+6Q^2,\qquad Q=b+b^{-1},\qquad
 E=-\frac{1}{24}+P^2,\qquad
 \bar E=-\frac{1}{24}+\Pbar^2.
 \label{eq:virasoro-parameters}
\end{equation}
In the semiclassical limit ($b\to 0$ with $bP$ and $b\bar{P}$ held fixed), we have $b\to 0$ and
$\frac{2\pi}{b}(P\pm\Pbar)\to 2\pi r_\pm/4G$, up to $\mathcal{O}(1)$ corrections, where $r_\pm$ are the outer and inner horizon radii of the BTZ black hole.
In this limit, simply expanding the hyperbolic sines into exponentials gives, up to an overall prefactor coming from loops,  
\begin{equation}
    e^{S_{0}(P,\bar{P})} 
    \sim 
    |\#| \exp\left[\frac{\mathcal{A}_{\textrm{out}}}{4G}\right]
    -|\#| 
    \exp\left[\frac{\mathcal{A}_{\textrm{in}}}{4G}\right]
    -
    |\#| 
    \exp\left[-\frac{\mathcal{A}_{\textrm{in}}}{4G}\right]
    +
   |\#| \exp\left[-\frac{\mathcal{A}_{\textrm{out}}}{4G}
    \right]\,,
    \label{eq:nonpert}
\end{equation}
with the distinct positive prefactors $|\#|$ coming from those $\mathcal{O}(1)$ additive corrections to $2\pi r_\pm/4G$.
It would be interesting to understand the connection between subleading saddles in the vacuum modular $S$-kernel and the full density of states $e^{S(h,\bar{h})}$ in more detail.
In any case, this result indicates that the inner horizon  does contribute as a subleading saddle in Cardy-type observables and 
supports our analysis in the previous section.

Finally, we comment on how the flat-space limit appears to isolate the inner horizon saddle identified above. In particular, the paper \cite{Bagchi:2012xr} that provided the original thermal interpretation of the FSC horizon 
employed an analog of~\eqref{eq:phasefactor}:
\begin{equation}
\begin{aligned}
    D(h_L,h_M)
    &=
    \int d\rho d\eta  e^{\frac{i \pi}{12 \eta^2}\left[c_L\eta -24  H_L \eta^3 -\rho( 24  H_M \eta^2  + c_M  )\right]}
    Z(\rho/\eta^2, -1/\eta)\,,
\end{aligned}
\label{eq:presaddle}
\end{equation}
where $\rho$ and $\eta$ are the Carrollian $\ell\to\infty$  limit of the CFT temperature  and chemical potential variables:
\begin{equation}
    \eta = \frac{1}{2}(\tau + \bar{\tau})
    \qquad \rho = \frac{\ell}{2}(\tau - \bar{\tau}) \,,
    \label{eq:etarho}
\end{equation}
and we defined
\begin{equation}
H_M=h_M-\frac{c_M}{24},\qquad
H_L=h_L-\frac{c_L}{24}\,.
\end{equation}
Unlike its AdS/CFT counterpart, this integral only has two saddles
\begin{equation}
    (\eta,\rho) = \pm\left(i\sqrt{\frac{c_M}{24 H_M}},\; i\frac{c_L H_M - c_M H_L}{2 H_M \sqrt{24 c_M H_M}}\right)\,,
\end{equation}
corresponding to the two inner horizon saddles in the AdS uplift.
Hence, taking the flat limit before performing the inverse-Laplace integral removes the divergent outer horizon saddles.
In future work, it would be interesting to find a better justification for why the flat limit can be taken inside the integral.

\section{Discussion}

In this work, we examined the flat limit of AdS$_3$/CFT$_2$, with a focus on holographic entanglement entropy proposals.
We first showed that ordinary RT surfaces  are ill-defined in the flat limit,
and we explained this from the boundary perspective by using a novel In\"on\"u-Wigner contraction to describe the Carrollian limit of local twist operator insertions.
In the process, we noticed that a change of conformal frame can lead to different BMS field theories being obtained in the Carrollian limit of the same CFT$_2$, 
which can then be used to explain why the swing entropy remains finite  while the RT entropy diverges.
We then turned to an investigation of  the swing surface in the AdS/CFT uplift.

As our first main result in this direction, we argued that bulk swing surfaces in AdS compute (half) the mutual information of nearly null intervals, in the vacuum state and a restricted family of other pure CFT states in which the entropy factorizes into left- and right-moving pieces.
We gave a bulk interpretation of this result using the scattering formalism of \cite{May:2019odp}.
As our second main result, we proved that the existence of a boundary dual for inner horizons in AdS/CFT implies  the existence of a boundary dual for swing surfaces in AdS/CFT.
This relies heavily on the work of \cite{Jiang:2017ecm}, but with some conceptual clarifications needed for the uplift. 
In both cases, the null rays defining the swing surface arise from ordinary UV cutoffs in the dual CFT. 
We also argued that the inner horizon contributes to CFT quantities through a subleading saddle in the Cardy formula, 
providing more motivation for the inner horizon quantity appearing in this construction.

In what follows, we summarize the main messages from the two approaches we have taken, and we highlight interesting future directions.

\paragraph{Operators in the Carrollian limit}
Our analysis exposes an unresolved question about the operator content of holographic CFT$_2$'s in the Carrollian limit.
In the standard conformal frame where Virasoro symmetries act as
$L_n=-z^{n+1}\partial_z$ and
$\bar{L}_n=-\bar{z}^{n+1}\partial_{\bar{z}}$ near $z=0=\bar{z}$,
we studied the Carrollian contraction of the conformal algebra to the BMS$_3$ algebra.
Under this limit, we found that local CFT primary operators with weights $h$ and $\bar{h}$ can become local BMS primary operators with  weights
\begin{equation}
    h_L = h+\bar{h}
    \qquad
    h_M= \frac{1}{\ell}(h-\bar{h})\,
    \label{eq:weights2}
\end{equation}
being ill-defined.
For instance, a local twist operator in the parent CFT has $h+\bar{h}\sim \ell$ 
and hence $h_L$ diverges in the Carrollian limit. 
Conversely, to obtain a BMS primary operator with finite BMS weights and $h_M\ne 0$, the CFT dimension $\Delta = h+\bar{h}$ must remain finite while the spin $s=h-\bar{h}$ grows as ${\cal O}(\ell)$.
This violates the local-operator unitary bound in the CFT, 
or equivalently forces one of the weights $h$ or $\bar{h}$ to become large and negative.

This discrepancy between CFT primaries and BMS primaries
--- much like the central-charge discrepancy that we focused on before ---
originates in the fact that the CFT state-operator map becomes singular in the Carrollian limit.
This raises the question of which objects in the parent CFT actually 
have a good flat limit. 
A fruitful approach in this direction is to consider the flat-space extrapolate dictionary for BMS primary operators proposed in \cite{Hijano:2017eii}.
In this dictionary, a boundary point $x$ on $\mathcal{I}^+$ determines a radial bulk null geodesic $\gamma_{x}$ in Minkowski spacetime, and one has the operator equation
\begin{equation}
\mathcal{O}_{h_L,h_M}(x)
= \int_{\gamma_x} d\lambda\;  \Phi_{h_L,h_M}(X(\lambda)) \,,
\end{equation}
where $\lambda$ is an affine parameter along $\gamma_{x}$.
The right-hand side of the above equation can be lifted to AdS/CFT using the formalism reviewed in section \ref{sec:basics}. In follow-up work we apply the HKLL dictionary \cite{Hamilton:2006az} to this integrated bulk field to understand what the corresponding boundary operator is, and find that it is related to certain non-local operators in CFT$_2$ built from lightray transforms \cite{Kravchuk:2018htv}.
In particular, the lightray transform swaps the large dimension for a large spin, as needed for finite BMS weights in~\eqref{eq:weights2}. 
We speculate that an appropriate truncation of the theory to those operators which remain well-defined in the Carrollian limit
can help us understand how to recover the correct flat-space central charges in the operator-frame limit.

\paragraph{Are swings covariant?}
An infamous feature of the swing surface proposal is that the two null rays defining the swing surface  --- called the ``ropes'' --- are traditionally defined in a coordinate-dependent way:
they are required to be \textit{radial} null rays which intersect the locus $r=0$ in Bondi gauge.
The constructions in sections~\ref{sec:lift} and \ref{sec:innerhor} shed light on this, because in both cases the radial prescription naturally arose from a choice of UV cutoffs in the parent CFT.

Specifically, in section \ref{sec:lift}, 
we connected swing surfaces in AdS$_3$ to boundary mutual informations that could further be related to a forward scattering process.
In this case, the two ropes arise as the limiting edges of the squeezed causal regions $J^+(c_1)\cap J^-(r_2)$ and $J^+(c_2) \cap J^-(r_1)$, respectively, where the relative positioning of the points $r_i$ and $c_i$ depend explicitly on the choice of UV cutoffs in the boundary theory. 
The radial prescription was thus tied to a particular symmetric prescription for 
the UV cutoffs, or equivalently the choice of conformal frame.
Similarly, in section \ref{sec:innerhor},
 we saw how UV regulators  
naturally extend into the bulk as complex geodesics,
first taking the form of a spacelike curve extending 
between the asymptotic boundary and the RT surface, 
and then as a timelike curve extending deep into the bulk beyond the RT surface, becoming null only as the UV regulator is removed.
Again, we found the resulting curve was radial only for a specific symmetric choice of UV cutoff prescription.
These results suggest that the failure of the swing to be covariant is to be viewed as a feature rather than a bug,
since it tracks the choice of UV cutoffs in the boundary theory.
In fact, a similar ``cutoff-covariance'' effect already occurs for RT surfaces in AdS; e.g.~see the treatment in \cite{Sorce:2019zce}.

We also note that this frame dependence echoes the frame dependence encountered in the symmetry contractions of section \ref{sec:basics}.
Hence, we can say that our work  points to a broader lesson whereby apparent ambiguities in flat holography often originate in a choice of frame in the parent CFT.

\paragraph{Swings for holographic scattering and sub-AdS scale physics}
In order to obtain a bulk dual for the mutual information in section \ref{sec:lift},
we used the ridge surface of \cite{May:2019odp}. 
However, beyond the vacuum setting, the ridge surface need not compute the mutual information directly.
On the other hand, we have shown that there exists a class of excited states dual to particular spinning defect geometries  in AdS$_3$ for which 
the swing surface computes a mutual information, up to some care with respect to the prescription for how to select among swing candidates. 
Hence it is natural to wonder whether our swing construction can be upgraded to yield an improved definition of the original ridge surface.
Recent work on bulk entanglement wedges~\cite{Arayath:2026rll}  suggests it might be productive to instead look for entropic quantities tied to the same bulk scattering region which however differ from the original ridge. 
Using this approach, and more generally, it would be interesting to  use the swing proposal as motivation to look for QI quantities on the boundary that capture sub-AdS regimes.

\paragraph{Inner horizon duals?}
Our main result in section \ref{sec:innerhor} was that the existence of an information-theoretic dual for inner horizons in AdS/CFT implies the existence of an information-theoretic dual for swing surfaces in AdS/CFT.
This then begs the question of what kind of duality we have in mind. 
We explored this in further detail in section~\ref{sec:subCardy}, where we showed that bulk inner horizon areas may appear in a subleading Cardy saddle of the boundary CFT$_2$. 
Our investigations  also seemed to point to the possibility of extracting some features of the inner horizon geometry from observables built from highly boosted intervals on the boundary. 

It would be interesting to connect our Cardy calculations with recent works investigating how inner horizon saddles arise in microcanonical gravitational path integrals \cite{Kruthoff:2024gxc,Caminiti:2026efx}. 
In particular,
our  observation in  \eqref{eq:nonpert} suggests that inner horizon saddles may play an important role in $3d$ quantum gravity, i.e.~beyond the JT settings considered previously.
Also, the fact that inner horizon saddles are complex \cite{Kruthoff:2024gxc,Caminiti:2026efx} is reminiscent of the complex continuation of the regulator ``rope'' geodesics through the outer horizon in section \ref{sec:modCHM}, suggesting a deeper connection between inner horizon saddles and swing surfaces.

It would also be interesting to understand the precise status of the associated ``negative-temperature'' saddle points in the CFT.
In particular, exact CFT$_2$ partition functions have a wall of singularities at infinite temperature, which apparently forbids continuation to negative-temperature saddle points.
A simple $3d$ toy model in which to investigate how to get around this issue would be the  end-of-the-world brane model of \cite{Jafferis:2025jle}.

\paragraph{From inner horizon to swing entropy}
From \eqref{nonpert0},  it is tempting to conclude that
swing surfaces compute exponentially subleading corrections to the entanglement entropy $S(\rho_A)$. 
However, this would be too quick.
This is because $S(\rho_A)$ is most naturally viewed as an entropy in the canonical ensemble 
rather than in the microcanonical ensemble.
To clarify the ensemble distinction, define the following conventions for the modular Hamiltonian $K_A$:
\begin{equation}
    \rho_A= e^{-K} = \frac{e^{-2\pi K_A}}{Z_A(2\pi)}
    \,, \qquad Z_A(\beta) \equiv \tr e^{-\beta K_A}\,.
\end{equation}
The von Neumann entropy is then the canonical quantity
\begin{equation}
    S(\rho_A) = \left(1-\beta\partial_\beta\right) \log Z_A(\beta)\bigg|_{\beta=2\pi}\,.
\end{equation}
By contrast, the fixed-modular-energy analogue of a microcanonical density of states is
\begin{equation}
    \Omega_A(k)\equiv \tr \left[\delta(K_A-k)\right]\,.
\end{equation}
Thus, the natural question is  whether the microcanonical observable $\Omega_A(k)$, not the canonical observable $S(\rho_A)$, receives a subleading saddle contribution from the swing length.
Making this idea precise requires developing a microcanonical formulation of subregion entropies in AdS/CFT.
An interesting future direction would therefore be to revisit the celebrated path-integral proof of the RT formula in \cite{Lewkowycz:2013nqa}
using microcanonical boundary conditions (see related work in \cite{Brown:1992bq,Marolf:2018ldl,Guo:2020roc,Baiguera:2024ffx}),
and ask whether subleading contributions from inner horizons and swing surfaces naturally appear.

Alternatively, in \cite{Caminiti:2026efx}, the
inner horizon geometry also appears as a negative-boundary-length saddle directly in the canonical ensemble.
This construction is suggestively related to the
spacelike modular flows studied in \cite{Jiang:2017ecm}, since those can be obtained  here by
reversing the sign of one chiral inverse temperature in
\eqref{newconf}. 
The conjecture then is that inner horizon areas compute von Neumann entropies in an unusual state with a negative temperature:
\begin{equation}
    S(\rho_{\beta_+, -|\beta_-|})
    =
    \frac{\textrm{Length}(\gamma_{\textrm{inner}})}{4G}\,
\end{equation}
which translates to
\begin{equation}
    S(\rho_{\textrm{spacelike}})
    =
    \frac{\textrm{Length}(\gamma_{\textrm{swing}})}{4G}\,,
\end{equation}
in the subregion setting.
A subtlety here is that the trace appearing in the von Neumann entropy diverges at negative modular temperature,  another manifestation of the negative temperature issue discussed above.

\paragraph{Other lessons for flat holography}
While we have encountered various lessons for AdS/CFT coming from flat holography, the main theme of these investigations has been to try to understand the limitations and successes of what we can import from AdS/CFT to AFS/BMSFT. 
In this vein, we note that  the swing surface is just one example of a more general picture where the ends of minimal surfaces are restricted in asymptotically flat spacetimes~\cite{Ghosh:2023kuy,Jorstad:2026jlg}. It would therefore be interesting to further explore the connections between the swing proposal and other attempts at flat-space holography.  Some references such as~\cite{Donnay:2022aba,Donnay:2022wvx,Donnay:2022sdg,Kulp:2024scx} lay out the kinematical relations between Carrollian~\cite{Duval:2014uva, Bagchi:2016bcd, Ciambelli:2018wre, Ciambelli:2018ojf, Bagchi:2019xfx,Bagchi:2019clu
} 
and celestial~\cite{ Strominger:2017zoo, Raclariu:2021zjz,Pasterski:2021rjz,Pasterski:2021raf} descriptions. 
In particular, it would be interesting to understand how the computations here, of the entanglement entropy of a subregion of a cut of null infinity, relate to earlier studies of entanglement across cuts of null infinity~\cite{Kapec:2016aqd, Chen:2023tvj,Chen:2024kuq}.
While the former is viewing null infinity as a limit of the timelike boundary of AdS, the latter is viewing it as the late-time limit of a bulk Cauchy slice. 
Related to the description of $\mathcal{J}^\pm$ as timelike-separated bands of the AdS boundary,
another direction to connect to is recent investigations of timelike entanglement entropy \cite{Doi:2023zaf,Heller:2024whi,Milekhin:2025ycm}. 
In particular, an interesting feature of the swing proposal is that the null ropes defining the swing connect the original interval $A$ to an antipodally oriented region $A'$ living in the past of $A$, and this has a similar structure to these timelike entanglement measures.
Overall, understanding how the swing proposal and BMSFT ``EE'' generalize to higher dimensions would help facilitate the comparison.

\section*{Acknowledgments}

It is a pleasure to thank 
Jan Albert, 
Arjun Bagchi, 
Glenn Barnich, 
Horacio Casini, 
Scott Collier, 
Luca Ciambelli, 
St\'ephane Detournay, 
Davide Gaiotto, 
Jaume Gomis, 
Daniel Grumiller, 
Jonathan Harper, 
Aidan Herderschee, 
Janet Hung, 
Eivind J\o rstad, 
Juan Maldacena,
Alex May, 
Dominik Neuenfeld, 
Jiaxin Qiao, 
Ana-Maria Raclariu, 
Max Riegler, 
Andrew Rolph, 
Joan Simon, 
Bruno Sivilotti,
Wei Song, 
Marija Tomaševi\'c, 
and Chris Waddell for valuable discussions, as well as Rob Myers for collaborating at many stages of this work. 
JC acknowledges the support from the Natural Sciences and Engineering Research Council of Canada through a Vanier Canada Graduate Scholarship [Funding Reference Number CGV-192707].  SP is supported by the Celestial Holography Initiative at the Perimeter Institute for Theoretical Physics and the Simons Collaboration on Celestial Holography.  Research at the Perimeter Institute is supported by the Government of Canada through the Department of Innovation, Science and Industry Canada and by the Province of Ontario through the Ministry of Colleges and Universities.

\appendix

\section{Revisiting BMS subalgebras of Virasoro}
\label{app:more-bms}
In  this appendix, we undertake a more systematic search for the different BMS$_3$ subalgebras of Vir$\times$Vir that one  can obtain using In\"on\"u-Wigner contractions. 
Our starting point is the ansatz
\begin{equation}
\badat{2}
    {\cal L}_n &= a_n L_n + \bar{a}_n \bar{L}_n - b_n L_{-n}  - \bar{b}_n \bar{L}_{-n}\, \\
    {\cal M}_n &=\frac{1}{\ell}( c_n L_n + \bar{c}_n \bar{L}_n - d_n L_{-n}  - \bar{d}_n \bar{L}_{-n})\,,
    \label{eq:ansatzzApp}
    \eadat
\end{equation}
where the coefficients are c-numbers that we will also take to be independent of $\ell$. Here $\ell^{-1}$ acts as a formal expansion parameter.  The scaling above then forces the supertranslations to commute $[{\cal M}_n,{\cal M}_m]=0$ when we are truncating to linear order in this parameter.\footnote{Note that while the vanishing commutator is enforced by a formal expansion in the algebraic picture, in the vector-field picture of section \ref{sec:vectors}, one must account for the simultaneous coordinate rescaling. Indeed, $\tau= u/\ell$ implies $\partial_\tau=\ell\partial_u$, compensating the explicit factor of $1/\ell$ in $\mathcal M_n$ and yielding finite, order-one vector fields. Their commutativity then follows from their specific $f_n(\phi)\partial_u$ form.}

To constrain the ansatz, we turn to the $[{\cal L}_n,{\cal L}_m]$ commutator and try to reproduce  \eqref{bms3}. We see that the right hand side should only contain modes with label $m+n$. 
This leaves us with two options: 
\begin{equation}
\begin{cases}
    {\cal L}_n = a_n L_n  - \bar{b}_n \bar{L}_{-n}\,, \\
    {\cal L}_n' = a_n L_n  + \bar{a}_n \bar{L}_n\,,
    \end{cases}
\end{equation}
up to the rather trivial possibility of exchanging the roles of $L_n$ and $\bar{L}_n$.
We will treat these two cases separately and so in particular we can rename $\bar{b}_n$ to $\bar{a}_n$ for simplicity.

In both cases, demanding that the BMS algebra is obeyed gives the recursive relations 
\begin{equation}
    a_{n+m} = a_n a_m  \qquad \bar{a}_{n+m}=\bar{a}_n\bar{a}_m\,,
\end{equation}
which are uniquely solved by
\begin{equation}
    a_n =a_1^n \equiv a^n
    \qquad
    \bar{a}_n =\bar{a}_1^n \equiv \bar{a}^n
    \,.
\end{equation}
That is, taking care of central terms also, we have
\begin{equation}
\begin{cases}
    {\cal L}_n = a^n L_n  - \bar{a}^n \bar{L}_{-n}\qquad 
    \qquad
    c_L = c-\bar{c}\,, \\ 
    {\cal L}_n' = a^n L_n  + \bar{a}^n \bar{L}_n\,,
    \qquad 
    \qquad
    c_L' = c+\bar{c}\, .
    \end{cases}
\end{equation} 
Note that we take $a$ and $\bar{a}$ to be nonzero.
These factors can then be removed using the automorphism $L_n\rightarrow a^nL_n$ and similarly $\bar L_n\mapsto \bar a^n \bar L_n$. Up to these automorphisms we then have
\begin{equation}
\begin{cases}
    {\cal L}_n = L_n  - \bar{L}_{-n}\qquad
    \qquad
    c_L = c-\bar{c}\,, \\ 
    {\cal L}_n' = L_n  + \bar{L}_n\,,
    \qquad
    \qquad
    c_L' = c+\bar{c}\, .
    \end{cases}
\end{equation}
Note that the manipulations in this appendix more readily follow whenever the CFT central charges have been chosen to scale with $\ell$ such that all obtained BMS central charges $c_L$, $c_M$, $c_L'$, $c_M'$ are $\mathcal{O}(\ell^0)$; in particular, we are not addressing the subtleties that arise for $c,\bar c\sim \ell$ in the manipulations here.

Now to constrain ${\cal M}_n$, we must demand that it satisfies the appropriate commutator with ${\cal L}_n$.
Doing so, we find 
\begin{equation}
\begin{cases}
    {\cal M}_n = \frac{1}{\ell}\left(\eta_1 (L_n+\bar L_{-n})+\eta_2 {\cal L}_n\right) \qquad
    \qquad
    &c_M = \frac{1}{\ell}(\eta_1(c+\bar{c})+\eta_2 c_L)\,, \\
    {\cal M}_n' =\frac{1}{\ell}\left(\eta_1 (L_n-\bar L_{n})+\eta_2 {\cal L}'_n\right) \qquad 
    \qquad
   & c_M' = \frac{1}{\ell}(\eta_1(c-\bar{c})+\eta_2 c_L') .
    \end{cases}
\end{equation}
with $\eta_i$ independent $\mathbb{C}$-numbers (but we take $\eta_1\ne 0$). However, now we can use that the limiting algebra is unchanged under 
\be
\begin{array}{ll}
&({\cal M}_n,c_M)\rightarrow (A {\cal M}_n, Ac_M),~~~~~~~~ A\neq 0,\\
&({\cal M}_n,c_M)\rightarrow ( {\cal M}_n+\frac{1}{\ell} B{\cal L}_n, c_M+\frac{1}{\ell}Bc_L),\\
\end{array}
\ee
to reduce this to the two canonical families we encountered in the main text. Namely,
\begin{equation}
\begin{array}{rll}
    {\cal L}_n &= L_n-\bar L_{-n} ~~~   \qquad & c_L=c-\bar c \\
    {\cal M}_n &= \frac{1}{\ell}( L_n+\bar L_{-n}) ~~~  \qquad    & c_M=\frac{1}{\ell}(c+\bar c),
    \label{eq:class1}
    \end{array}
\end{equation}
and
\begin{equation}
\begin{array}{rll}
    {\cal L}'_n &= L_n+\bar L_{n} ~~~   \qquad & c_L'=c+\bar c \\
    {\cal M}'_n &= \frac{1}{\ell}( L_n-\bar L_{n}) ~~~  \qquad    & c_M'=\frac{1}{\ell}(c-\bar c) .
    \label{eq:class2}
    \end{array}
\end{equation}
In particular, we see that the state-frame and operator-frame contractions are in classes~\eqref{eq:class1} and~\eqref{eq:class2}, respectively.

\section{Swings in the spinning defect geometry}
\label{app:spsm_check}

In this appendix, we compute candidate swing surfaces in the spinning defect geometry.
We will be interested in the candidate swing surfaces associated with the boundary interval $A$ with endpoints $(\pm \Delta \tau/2, \pm \Delta\phi/2)$.
To find the first swing candidate, our strategy will be to solve for the locus at which the past lightsheets emanating from $\partial A_1$ and $\partial A_2$ intersect,
and then  implement the radial null-rope cutoffs.

Our starting point is the spinning defect metric in \eqref{eq:defect-metric-bpm}:
\begin{equation}
ds^2 = 
-
\frac{(r^2+b_+^2)(r^2+b_-^2)}{r^2} d\tau^2
+
\frac{\ell^2 r^2 dr^2}{(r^2+b_+^2)(r^2+b_-^2)} 
+ r^2 \left(d\phi - \frac{b_+ b_-}{r^2}  d\tau\right)^2.
\label{eq:defect-metric-bpm-app}
\end{equation}
Since we are interested in generic configurations and not the flat limit, we will set $\ell=1$ for notational simplicity in what follows. Also for notational simplicity, we introduce the following shorthand notations:
\begin{equation}
    \begin{cases}
        D \equiv b_+^2-b_-^2\\
        C \equiv \cos(\Delta T)-\cos(\Delta \Phi)
    \end{cases}
    \qquad 
    \begin{cases}
    \alpha =\Delta T \equiv  b_+ \Delta \tau + b_-\Delta \phi\\
    \beta =\Delta \Phi \equiv
     b_-\Delta \tau  +  b_+\Delta \phi
     \end{cases}
     \qquad
     \begin{cases}
    \Delta Y^+ =\frac{\Delta T + \Delta \Phi}{2}
     \\
     \Delta Y^-=\frac{\Delta \Phi-\Delta T }{2}\,.
     \end{cases}
\end{equation}
Then, we perform a sequence of coordinate changes.
First, we introduce coordinates
\begin{equation}
    T=\tau b_++ \phi b_- \qquad 
    \Phi = \tau b_- + \phi b_+
    \qquad
    R^2 = \frac{r^2 + b_-^2}{D}\,,
\end{equation}
in which the metric is locally that of global AdS$_3$:
\begin{equation}
    ds^2 
    =
    -(R^2+1)dT^2 
    +\frac{dR^2}{R^2+1}+R^2 d\Phi^2\,.
\end{equation}
Then, we introduce embedding coordinates 
\begin{equation}
    (X_{-1},X_0, X_1,X_2)
    =
    (\sqrt{R^2+1}\cos T,
    \sqrt{R^2+1}\sin T,
    R\cos \Phi,
    R\sin\Phi)\,,
    \label{mbed}
\end{equation}
within the $(-1,-1,1,1)$ signature flat space $\mathbb{R}^{2,2}$. The metric is then just the flat-space line element restricted to the unit-radius hyperboloid $-X_{-1}^2 -X_{0}^2+X_1^2+X_2^2=-1$.

Now, in these embedding coordinates, we can represent a point $(T_*, \Phi_*)$ at the asymptotic boundary of AdS by a null vector
\begin{equation}
    P(T_*, \Phi_*)
    =
    (\cos T_*, \sin T_*, \cos \Phi_*, \sin \Phi_*)\,.
    \label{mbedbdy}
\end{equation}
Its associated lightcone is the space of points $X$ such that
\begin{equation}
     P(T_*, \Phi_*)\cdot X = 0\,.
\end{equation}
Hence, to compute the intersection of the past lightcones emanating from $\partial A$, 
we are interested in the locus
\begin{equation}
    P\left(\frac{\Delta T}{2}, \frac{\Delta \Phi}{2}\right)\cdot X = 0,~~
    P\left(-\frac{\Delta T}{2}, -\frac{\Delta \Phi}{2}\right)\cdot X =0\,.
    \label{pstars}
\end{equation}
Solving this system of equations as well as the constraint $X^2=-1$, we find the following curve:
\begin{equation}
    X(\lambda) = \frac{1}{\sqrt{2C}}\left[e^\lambda 
    P\left(-\frac{\Delta \Phi}{2},-\frac{\Delta T}{2}\right) 
    + e^{-\lambda} 
    P\left(-\pi+\frac{\Delta \Phi}{2},-\pi+\frac{\Delta T}{2}\right) \right]\,.
    \label{eq:xlambda}
\end{equation}
We have parameterized this curve to obey $\dot{X}^2=1$,
so that $\lambda$ parameterizes the proper length.
We have also chosen the overall sign to ensure that this curve lies in the intersection of the past lightcones (rather than the future ones).

Our next step is to study an individual null ray $X(s)$ which emanates from a general boundary point $P_*$.
We claim that  $X(s)$ has the following form:
\begin{equation}
    X(s) = X_0 + sP_*\,,
    \label{nullray}
\end{equation}
for some $X_0$ in the past lightcone of $P_*$, satisfying $P_*\cdot X_0=0$.
To see this, simply note that the curve $X(s)$ as defined above is null,
obeys the embedding equation $X(s)^2=-1$ everywhere,
and lies in the lightcone of $P_*$ (i.e.  $P_*\cdot X(s)=0$).
The main utility of this formula is to see that the tangent of 
the curve $X(s)$ at any point is  $P_*$.

We now examine which null rays are radial null rays.
By definition, radial null rays $X(s)$ have vanishing angular momentum $p_\phi$ with respect to the angular Killing field $\partial_\phi$ in the original spinning defect coordinates in \eqref{eq:defect-metric-bpm-app}.
To translate this statement to embedding space,
first note that symmetries of AdS can conveniently be represented as linear operators on embedding space.
In particular, using \eqref{mbed}, we can derive the identities
\begin{equation}
    \partial_T X_a = 
    \begin{pmatrix}
        0 & -1 &  0 & 0\\
        1 & 0 &  0 & 0\\
        0& 0 &  0 & 0\\
        0& 0 &  0 & 0
    \end{pmatrix}
     \begin{pmatrix}
        X_{-1}\\
        X_0\\
        X_1\\
        X_2
    \end{pmatrix}
    \qquad
    \partial_\Phi X_a = 
    \begin{pmatrix}
        0 & 0 &  0 & 0\\
        0 & 0 &  0 & 0\\
        0& 0 &  0 & -1 \\
        0& 0 &  1 & 0
    \end{pmatrix}
     \begin{pmatrix}
        X_{-1}\\
        X_0\\
        X_1\\
        X_2
    \end{pmatrix}
\end{equation}
which give us matrices $K_T$ and $K_\Phi$  representing the time-translation  and rotation symmetries of global AdS$_3$.
To represent the angular symmetry of the spinning defect metric,
we can use the chain rule to write
$\partial_\phi = b_-\partial_T+b_+ \partial_\Phi$, 
and hence
\begin{equation}
    \partial_\phi X_a = K_\phi X_a
    \label{linearize}
\end{equation}
with $K_\phi= b_- K_T + b_+ K_\Phi$.
Note that 
these matrices can also be used to implement symmetry actions on the boundary.
In particular, using \eqref{mbedbdy},
we have the identities
$K_T P(T_*,\Phi_*)=\partial_{T_*}P(T_*,\Phi_*)$
and 
$K_\Phi P(T_*,\Phi_*)=\partial_{\Phi_*}P(T_*,\Phi_*)$,
where the derivatives are now taken with respect to boundary points $T_*$ and $\Phi_*$.

Now we are ready to revisit the angular momenta $p_\phi$ of our null rays.
The Killing vector field associated with this conserved charge is
\begin{equation}
    \partial_\phi
    =
    \frac{\partial X_a}{\partial \phi}\partial_{X_a}
    \,,
\end{equation}
and can be identified with the  vector $\partial_\phi X=K_\phi X$ in $\mathbb{R}^{2,2}$.
Hence, the angular momentum of a null ray \eqref{nullray} through the point $X$ reads simply
\begin{equation}
    p_\phi = K_\phi X \cdot P_*\,.
\end{equation}
For the swing surface, we are interested in the particular choice where $X=X(\lambda)$ are on the curve in \eqref{eq:xlambda} while $P_*$ describes the endpoints of $A$.
Hence, for $P_*=\partial A_i$ with $i=1,2$, we must set to zero
\begin{equation}
    P_* \cdot K_\phi X(\lambda)
    =
   \frac{e^\lambda}{\sqrt{2C}}
    P_*\cdot K_\phi P\left(-\frac{\Delta \Phi}{2},-\frac{\Delta T}{2}\right) 
    + \frac{e^{-\lambda} }{\sqrt{2C}}
    P_*\cdot K_\phi P\left(-\pi+\frac{\Delta \Phi}{2},-\pi+\frac{\Delta T}{2}\right) \,,
\end{equation}
to solve for the hit points $X(\lambda_i)$.

Solving these conditions for $\lambda_1$ and $\lambda_2$,  using the comment below \eqref{linearize}, yields
\begin{equation}
       e^{2\lambda_1} 
   =
   \frac{ (b_+ - b_-) \sin \Delta Y^+}{(b_+ + b_-) \sin \Delta Y^-}=e^{-2\lambda_2}
\end{equation}
and consequently the length of the swing candidate $|\lambda_1-\lambda_2|$ is simply given by $2|\lambda_1|$, or equivalently
\begin{equation}
    \textrm{Length}= \log \frac{(b_+ - b_-) \sin \Delta Y^+}{(b_+ + b_-) \sin \Delta Y^- }
    =
    \log \frac{(b_+ - b_-) \sin \frac{\Delta T + \Delta \Phi}{2}}{(b_+ + b_-) \sin \frac{\Delta \Phi- \Delta T}{2}}\,,
\end{equation}
which  is \eqref{eq:entropy-diff-spinning-defect} in the main text, after also restoring factors of $\ell$. 
We suppress absolute values here and in \eqref{eq:entropy-diff-spinning-defect}.
The second candidate is obtained by sending $\Delta\phi\rightarrow 2\pi-\Delta\phi$ and $\Delta \tau\mapsto -\Delta \tau$.

\bibliographystyle{utphys}
\bibliography{references}

\end{document}